\documentclass[9pt]{article}

\usepackage{latexsym,amsmath,amscd,amssymb,amsthm,graphics}
\usepackage{subcaption}
\usepackage{makecell}
\usepackage[toc, page]{appendix}

\usepackage{enumerate,todonotes}
\usepackage[margin=2cm]{geometry}
\usepackage{graphicx}
\usepackage[colorlinks]{hyperref}
\usepackage{cleveref}
\usepackage{url}
\usepackage{framed, comment}
\usepackage[all]{xy}
\usepackage{epstopdf}
\usepackage{float}

\newtheorem{theorem}{Theorem}[section]
\newtheorem*{theorem*}{Theorem}

\newtheorem{remark}[theorem]{Remark}
\theoremstyle{definition}
\newtheorem{method}{Method}[section]
\newtheorem{ensemble method}{Ensemble Method}[section]
\newtheorem{example}{Example}[section]

\def\bx{\boldsymbol{x}}

\makeatother\def\b{\boldsymbol}

\begin{document}

\title{Stochastic fluids with transport noise: Approximating diffusion from data using SVD and ensemble forecast back-propagation}

\author{James Woodfield \footnote{Department of Mathematics, Imperial College, London SW7 2AZ, UK. }}

\date{\today}

\maketitle

\makeatother
\begin{abstract}{
We introduce and test methods for the calibration of the diffusion term in Stochastic Partial Differential Equations (SPDEs) describing fluids. We take two approaches, one uses ideas from the singular value decomposition and the Biot-Savart law. The other backpropagates through an ensemble forecast, with respect to diffusion parameters, to minimise a probabilistic ensemble forecasting metric. We describe the approaches in the specific context of solutions to SPDEs describing the evolution of fluid particles, sometimes called inviscid vortex methods. The methods are tested in an idealised setting in which the reference data is a known realisation of the parameterised SPDE, and also using a forecast verification metric known as the Continuous Rank Probability Score (CRPS). }
\end{abstract}

\section{Introduction}\label{sec:Introduction}

\subsection{History and motivation}\label{sec:history and motivation}

Motivated by the need to model the effect of viscosity not present in the inviscid vortex method, Chorin \cite{chorin1973numerical} proposed a constant (Itô) noise in the particle trajectory map. Chorin's stochastic parameterisation represented the diffusion effect present in the corresponding Fokker-Plank equation (the deterministic Navier-Stokes equation). Numerical methodology based on the idea of a stochastic particle trajectory map was later proven convergent in \cite{long1988convergence}, and are sometimes called computational vortex methods \cite{majda2002vorticity}.

Computational vortex methods are numerical methods based on tracking the particle trajectories of a finite number of discrete points of (potential) vorticity, making use of the Biot-Savart Law to both close the system and define the velocity elsewhere. Typically, one requires the approximation of an integral, the regularisation of a kernel, and the closure as a collocation method \cite{majda2002vorticity}. Possible advantages to computational vortex methods include not needing to allocate computational resources to regions with little or no vorticity, the absence of a pressure solve, little to no numerical viscosity, a less stringent timestep requirement, and access to the velocity globally.

More recently, in 2015, Holm proposed a different type of stochastic parameterisation of the particle trajectory map \cite{holm2015variational}, rather than modelling diffusion, the introduced stochastic parameterisation aims at representing uncertainty associated with additional transport. In this setting a family of spatially dependent vector fields $\lbrace\b \xi_{p}(\b x)\rbrace_{p=1}^{P}$ act (stochastically) on the particle trajectory map. The basis of vector fields are integrated (in the Stratonovich sense) against a $P$ dimensional Brownian motion, as to remain consistent with the variational principle, preserve Kelvin's theorem, and preserve infinite integral quantities known as Casimir's, such ideas are presented in \cite{arnold1966geometrie,holm2015variational,holm1998euler}. In practice one still requires methods for estimating the vector-fields $\lbrace\b \xi_{p}(\b x)\rbrace_{p=1}^{P}$, as to take into account the uncertainty associated with unresolved or unrepresented transport, doing so is the problem we tackle in this paper. This task will colloquially be described as calibration or as an offline batch data assimilation technique, the aim is to present a calibrated stochastic forward model capable of producing an ensemble which represents statistics of the data, or more generally the statistics of the hidden distribution from which the data was sampled. 

In Cotter et al., 2019 \cite{cotter2019numerically} and Crisan et al., 2023 \cite{crisan2023noise}, vector fields are calibrated from weather station positional data using an SVD/PCA/EOF decomposition of a Data-Anomaly-Matrix (DAM) formed in the context of stochastic coarsegraining of a high-resolution deterministic model, see also \cite{resseguier2021new,resseguier2020data} for application and variants of this methodology to other stochastic fluid models. In these works \cite{cotter2019numerically,resseguier2020data,resseguier2021new,crisan2023noise} stochastic parameterisation of the coarse-graining operator have been proposed in the context of stochastic model reduction. We instead consider parameterisation between a reference dataset and the proposed stochastic forward model, including when the data arises as a realisation of another stochastic model, not necessarily the proposed stochastic forward model. In this work we use a similar truncated SVD approach to calibrating a basis from weather-station data, however amongst other minor details we differ in the construction of the data anomaly matrix, the resulting calibrated proposed stochastic equation, and the numerical methods used. We test the new calibration technique using a twin experiment framework where the reference data is a known parameterised SDE/SPDE realisation (whose parameters are known). We also test in an extended twin experiment in which the distribution from which the data is sampled is known, and more SPDE/SDE realisations are generated for hidden testing datasets. 

We also introduce preliminary results regarding, a loss-based approach to the calibration problem. The motivation for another calibration method stems from the need for the calibration/assimilation of other types of data such as drifter data or simply state-valued data in the forward model, and is motivated by trying to alleviate the expensive interpolation from weather-station cost in the forward ensemble model.

\subsection{Outline}\label{sec:outline}
\begin{enumerate}
    \item \Cref{sec:Introduction} contained the motivation and history.
    \item \Cref{sec: Governing Equations and Numerical Methods} contains a review of the stochastic fluid modelling assumptions (\cref{sec:Governing Equations}) and the 2D computational vortex methods of interest (\cref{sec: Numerical methods}). 
    \item \Cref{sec:Calibration methodology} introduces several approaches to calibration.
    \begin{enumerate}
        \item \Cref{sec:svd and ml  Calibration methodology} contains the two proposed calibration methodologies. The first calibration approach focuses on the SVD decomposition of a space-time data anomaly matrix. In the second calibration approach, we treat an entire forecast ensemble method (over a space-time window) as a function, and a Continuous Ranked Probability Score estimator is used as a loss. The loss is minimised with respect to the parameterised basis of vector fields. 
        \item  \Cref{sec: ensemble methods} contains a list of the different ensemble forwards models that we will test, including benchmark ensembles.
    \end{enumerate}
    \item \Cref{sec:experiments} contains the details and results of several numerical experiments. 
    \begin{enumerate}
        \item We describe a twin experiment in which the reference data comes from a parameterised SDE approximation of an SPDE, with a fixed known single basis and a single known Brownian motion. 
        \item We describe a twin experiment in which the data comes from a parameterised SDE approximation of an SPDE, with 5 fixed known single basis functions and 5 i.i.d. Brownian motion realisations.
        \item We describe a twin experiment in which the synthetic data is generated from a realisation of a parameterised SDE approximation of an SPDE with an additional Ito-Stratonovich drift.
        \item We describe a twin experiment in which the synthetic data is generated as a realisation of an SDE system with a larger physical drift term representing the effect of unrepresented dynamics. 
        \item 
        We finally present results comparing Continuous Rank Probability Score (CRPS) and relative skill scores (CRPSS) for various calibration techniques proposed in this paper against some proposed benchmarks. 
    \end{enumerate}
    \item \Cref{sec: conclusion} concludes and summarises key results and preliminarily discusses the application of this calibration methodology to less idealised data.  
\end{enumerate}

\section{Governing Equations and Numerical Method}\label{sec: Governing Equations and Numerical Methods}
\subsection{Governing equations}\label{sec:Governing Equations}
Various stochastic parameterisations of the particle-trajectory mapping in fluid mechanics have been proposed \cite{chorin1973numerical,holm2015variational,crisan2022variational}. In this work, we are interested in the two-dimensional case where the initial label $\b X = (X,Y)^T \in \mathbb{R}^2$ is evolved to current configuration $\b x= (x,y)^T$ by the parameterised Stratonovich stochastic ordinary differential equation 
\begin{align}
\b x( \b X,t) = \b x(\b X,0) + \int_{0}^{t} \b u(\b x(\b X,s),s)d s + \sum_{p=1}^{P}\int_{0}^{t}  \theta_p \b \xi_{p}(\b x(\b X,s)) \circ d W^p(s); \quad \b x(\b X,0) = \b X .\label{eq:intro particle traj map}
\end{align}
Where $\b u(\b x,t) $ is the drift velocity, and $\lbrace \theta_p \b \xi_{p}(\b x)\rbrace_{p=1}^{P}$, is a set of velocity fields associated with the stochastic component of the flow. $\circ dW^p$ denotes Stratonovich integration against the $p$-th component of a $P$-dimensional Brownian motion, (see \cite{karatzas2012brownian}). Each vector field basis is multiplied by a parameter value $\theta_p \in \mathbb{R}$. We assume that the drift stream function $\psi$ is related to vorticity $\omega$ by a yet specified differential relationship, solvable by convolving the Green's functions against the vorticity as follows 
\begin{align}
\psi(\b x,t)= \int_{\mathbb{R}^2} G(\b x-\b x') \omega( \b x') d \b x'. \label{eq:stream function}
\end{align}
and that the negative skew gradient $\b u =-\nabla^{\perp}\psi$ relates the velocity $\b u$ to the vorticity $\omega$ by another convolution against the kernel $K$
\begin{align}
 \b u(\b x,t)= \int_{\mathbb{R}^2}  K(\b x-\b x') \omega( \b x') d\b x', \label{eq:biot savart}
\end{align}
this relationship is known as the Biot-Savart law, and when substituted into \cref{eq:intro particle traj map} describes the particle trajectory map integrodifferentially. Many fluid models have such a formulation, some important examples are discussed in the appendix \cref{example: 2d Euler alpha,example: 2d QGSW,example: 2d SQG}, and the ones used explicitly in this paper are discussed below (\cref{example: 2d Euler,example: regularised euler}).  To completely define the above infinite dimensional SDE system, (equivalent to the the solution of an SPDE fluid model) one typically defines an initial vorticity field $\omega_0$, and notes that this quantity is invariant along solution trajectories. See \cite{diamantakis2024evy} for additional information about deriving such a model from a variational principle.

\begin{example}[2D Euler]\label{example: 2d Euler}
    Euler on $\mathbb{R}^2$ has the following differential relationship between the drift stream function and vorticity, $\psi = 	(-\Delta)^{-1}\omega$, with Green's function given by $G(\b x)=-(2\pi)^{-1}\log(||\bx||_2)$, and kernel by $K(\b x) =(2\pi)^{-1} \bx^{\perp} ||\b x||^{-2}$. The corresponding SPDE is a stochastic version of Euler's equation given below in vorticity form
\begin{align}
    d \omega_t + (\b u \cdot \nabla )\omega_t dt + \sum_{p=1}^{P} (\theta_p \b \xi_{p}\cdot \nabla )\omega_t \circ dW^{p}, \quad \b u = -\nabla^{\perp} \psi, \quad \psi = (-\Delta)^{-1}\omega,
    \label{eq:2d salt euler}
\end{align}
where $\omega = \operatorname{curl}(\b u)$, is the vorticity. 
\end{example}

\begin{example}[Regularised Euler]\label{example: regularised euler}
It is often the case that the kernel $K$ possesses a singularity (at $\b x = 0$), making numerical methods approximating the velocity field from a delta function initial condition ansatz and the Biot-Savart kernel \cref{eq:biot savart} inaccurate \footnote{In particular, Beale and Majda 1985 \cite{beale1985high}, show that the point vortex method for the Euler equation has poorer convergence properties as the number of points increases, as compared with methods employing a regularised kernel. Furthermore, they show point vortex methods have a larger error when evaluating velocities not on point vortex trajectories, as compared with their vortex blob counterparts.}. The kernel $K$ is instead typically regularised by component-wise convolution with a parameterised mollifier function $\phi_{\delta}$, $\delta\in \mathbb{R}^{>0}$, such that the resulting kernel \begin{align}
    K_{\delta} = K\ast \phi_{\delta} = \int_{\mathbb{R}^2}K(\b x-\b y)\phi_{\delta}(\b y)dy,
\end{align} is desingularised, and the regularised Biot-Savart law is given by
\begin{align}
 \b u^{\delta}(\b x,t)= \int_{\mathbb{R}^2}  K_{\delta}(\b x-\b x') \omega( \b x') d\b x' = \int_{\mathbb{R}^2}  K(\b x-\b x') \omega_{\delta}( \b x') d\b x'. \label{eq:regularised biot savart}
\end{align}
Where in the last line, the associative property of convolution has been used to give interpretation as a regularised Euler vorticity $\omega_{\delta} = \omega\ast \phi_{\delta}$. 
\end{example}
In this work, we are interested in a manner of determining sensible proposals for $\lbrace\theta_p \b \xi_{p} \rbrace_{p=1}^{P}$ from data, such that a forward model numerical method can produce a probabilistic forecast with skill. More specifically in this work, we are interested in parameterising the difference between reference data and the forward model, a Stochastic Advection by Lie Transport (SALT) inviscid vortex dynamics solver. We will test on idealised reference data arising from realisations of similar known stochastic forward models. This is an idealised setting in which we can test the calibration methodologies, without modelling error concerns. However, the application of the methodology can be speculated to be applicable in other modelling scenarios such as stochastic model reduction as in \cite{cotter2019numerically,resseguier2020data,resseguier2021new,crisan2023noise}. We will also do testing in the setting in which there is a known modelling discrepancy. Namely, we suppose that there exists an additional constant drift velocity between the forward model and the data. One example of such a model discrepancy would be interpreting the stochastic integration in a different setting, i.e. Itô-Stratonovich, or Wong-Zakai anomaly type drift \cite{diamantakis2024evy}. Another likely motivation is the assumption that in a time-averaged scenario, models simply differ by a drift from real-world data.

\subsection{Numerical method}\label{sec: Numerical methods}

Point vortex methods model the initial vorticity by a field whose vorticity is concentrated at a finite sum of delta functions whose strength is denoted $\Gamma_i\in \mathbb{R}$ as follows
\begin{align}
    \omega(\b x) = \sum_{ i}\Gamma_{ i}\delta(\b x -\b x_i).
\end{align} If the vorticity is assumed a finite sum of delta functions, using a regularised kernel $K_{\delta}$, is equivalent to approximating the vorticity with ``vortex-blobs" with finite width $\omega(\b x) = \sum_{\b i}\Gamma_{\b i}\phi_{\delta}(\b x -\b x_i)$, $\phi_{\delta} = \phi_{\delta} \ast \delta $ and using the unregularised Euler kernel $K$. The mollifier $\phi_{\delta}$ used in vortex blob regularisation's are typically constructed with specific smoothness and moment boundedness properties (pg227)\cite{majda2002vorticity} (pg190)\cite{beale1985high}, required for convergence and stability estimates (sec 6.4 and sec 6.6 \cite{majda2002vorticity}). In this work, we consider inviscid vortex methods, which approximate the regularised stochastic integrodifferential equation for 2d Euler on $\mathbb{R}^2$, and essentially use a stochastic version of the deterministic discretisation strategy proposed in \cite{beale1985high}, outlined below.

Let the multi-index $\b i = (i_1,i_2)$, belong to a finite index set $\b \wedge_0^{\b i}$ spanning (labelling) the dynamically evolving points $\b x_{\b i} = (x_{(i_{1},i_{2})},y_{(i_{1},i_{2})})$ with non zero initial vorticity, these points will initially be defined on a Cartesian mesh in $[x_{\min},x_{\max}]\times [y_{\min},y_{\max}]$ with uniform width and height $h_x,h_y$. One assumes that the deterministic part of the regularised vorticity field $\omega_{\delta}(\b x,t)$, velocity field $\b u_{\delta}(\b x,t)$ and stream function $\psi_{\delta}(\b x,t)$ can be reconstructed globally on $\mathbb{R}^2$, $\forall t \in [0,T]$ in the following way  
\begin{align}
\omega^{\delta}(\b x,t) &= \sum_{\b i\in \wedge^{\b i}_0 } \phi_{\delta}(\b x - \b x_{\b i}(t)) \omega_0(\b X_{\b i})h_x h_y,
\label{eq: anzatz}\\
\b u^{\delta}(\b x, t) &= \sum_{\b i\in \wedge^{\b i}_0 } K_{\delta}(\b x - \b x_{\b i}(t)) \omega_0(\b X_{\b i})h_x h_y, \label{eq: global biot savart}\\
\psi^{\delta}(\b x, t) &= \sum_{\b i\in \wedge^{\b i}_0 } G_{\delta}(\b x - \b x_{\b i}(t))\omega_0(\b X_{\b i})h_x h_y, \label{eq:global stream}
\end{align}
from the finite set of evolving points $\b x_{\b i}(t)\in \wedge_0(t)$. Where $G_{\delta}=G \ast \phi_{\delta}$, denotes the convolution of the Green's function with the mollifier, $K_{\delta}=K \ast \phi_{\delta}$ denotes the convolution of the Euler kernel with the mollifier, and in \cref{eq: anzatz} the molifier convolves a delta function to define the vortex blob function $ \phi_{\delta}=\phi_{\delta} \ast \delta$ at the positions $\b x_{\b i}$. Noting that (potential)vorticity is preserved along solution trajectories $\omega_0(\b X_i)= \omega(\b x_{i}(t))$, it is possible to interpret \cref{eq: global biot savart,eq:global stream}, as discretisations of the convolutions in \cref{eq:stream function,eq:biot savart}, with either vortex blob initial conditions, or with regularised convolutions.

Numerically, in practice \cref{eq: anzatz,eq: global biot savart,eq:global stream} are not evaluated globally, but will be evaluated at fixed weather-station positions for data denoted $\b x_{\b d}\in \wedge_{\b d}$, and moving vortex positions $\b x_{\b i}(t) \in \wedge_0(t)$.

Upon appropriate vectorisation of the initial mesh $\wedge_0$, and identification of the ``point" vortex strength $\Gamma_{\b i} = \omega_0(\b X_{\b i})h_x h_y, \quad 
\forall \b i\in \wedge^{\b i}_{0},$ one can use \cref{eq: global biot savart} to close the system as a finite-dimensional system,
\begin{align}
&\b x_{\b i}(t) = \b x_{\b i}(0) + \int_{0}^{t} \b u^{\delta}(\b x_{\b i}(s),s)ds + \sum_{p=1}^{P}\int_{0}^{t}\theta_p \b \xi_{p}(\b x_{\b i}(s))\circ dW^{p}_s, \quad \forall \b i \in  \wedge_{0} ,\label{eq: finite dimensional particle trajectory regularised} \\
&\b u^{\delta}(\b x_{\b i}(s),s) = \sum_{\b j \in \wedge^{\b i}_0,\b j\neq \b i } \Gamma_{\b j} K_{\delta}( \b x_{\b i}(s) - \b x_{\b j}(s)), 
\end{align}
where each vortex does not self-induce a velocity.

Various mollifiers can be used in inviscid vortex methods. For the simulation of the Euler equation (\cref{example: 2d Euler}),
Rosenhead \cite{rosenhead1931formation}, Krasney \cite{krasny1986study} and Chorin \cite{chorin1973numerical}, all introduced mathematical equivalents to mollification of the Euler kernel, preventing division by zero and cutting off the singularity. In 1979 Hald \cite{hald1979convergence} proved second order convergence for 2D deterministic vortex methods when using a specific mollifier over an arbitrary time interval. However, the specific form of mollifier (compact locally three times differentiable) required the regularisation parameter $\delta$ to be larger than the mesh spacing $h$. Beale and Majda in \cite{beale1982vortex} introduce smoother mollifiers allowing smaller regularisation parameter $\delta = O(h)$ and proved arbitrary order convergence. In \cite{beale1985high} Beale and Majda introduce convenient additional explicit higher order kernels, we adopt one such family of mollifiers in this work, and the effect on the the Biot-Savart kernel can be described in the following manner,  
\begin{align}
(u^{\delta}(x,y),v^{\delta}(x,y))^{T}  = \sum_{\b i \in \wedge_{0}} \frac{\Gamma_{\b i }(-(y-y_{\b i}),(x-x_{\b i}))^T}{2\pi || \b x - \b x_{\b i }||_2^2} (1 - L_{p}(|| \b x - \b x_{\b i }||_2^2 / \delta^2 ) \exp(-|| \b x - \b x_{\b i }||_2^2 /\delta^2)),
\label{eq: regularised biot savart}
\end{align}
where $L_{p}$ is the p-th order Laguerre polynomial, and can be found in \cite{beale1985high}. This scheme (in the deterministic setting) has been shown to have the property that if $\delta = h^{q}$ for $q \in(0,1)$, the order of convergence to the solution of the Euler Equation is given by $O(h^{(2p+2)q})$ see \cite{beale1985high} (or sec 6\cite{majda2002vorticity}).

To deal with the stochastic Stratonovich term, we discretise in time with the stochastic generalisation of the SSP33 scheme of Shu and Osher (can be found in \cite{shu1988efficient}), where the forward Euler scheme is replaced with Euler Maruyama scheme in the Shu Osher representation. This time-stepping is applied to \cref{eq: finite dimensional particle trajectory regularised}, the scheme is weak order 1, strong order 0.5, as can be found by Taylor expanding (see \cite{ruemelin1982numerical} for the strict generalisation of this result) and (\cite{kloeden1992stochastic} for definitions of convergence).

\section{Calibration Methodology}\label{sec:Calibration methodology}

\subsection{Procedures and methodology: in the estimation of basis functions and parameters}\label{sec:svd and ml  Calibration methodology}

This section details two methods used in the estimation of $\lbrace\theta_p \rbrace_{p=1}^{P}$, the recovery of basis functions $ \lbrace \sigma_p \tilde{\b \xi}\rbrace_{p=1}^{P}$, the recovery of the time mean difference $\overline{\b v}^r $ and the recovery of paths $\lbrace \Delta W^{r}_{p} \rbrace_{p=1}^{P}$. 




The first method takes inspiration from the coarse-graining parameterisation approaches taken in \cite{cotter2019numerically,crisan2023noise} in the use of the SVD. However, the aim of the model here is to parameterise the difference from the reference data and the forward model. This is done by using the Biot-Savart kernel to ``access" the drift component of the velocity directly in the creation of a data anomaly matrix. In practice the details of the algorithm are given below.


\begin{method}[TSVD$\circ$WSD] Truncated Singular Value Decomposition of weather-station data. \label{method:TSDVWSD}
\begin{enumerate}
\item Data Collection; We assume that over the discrete time interval $\mathcal{T}:= \lbrace t_{n}\rbrace_{n=0}^{n_{t}}$, we have recorded a velocity field $\b u^{m}_{\b d} = \b u^{m}(\b x_{\b d
})$, measured at the fixed weather station positions $\b x_{\b d} = (x_d,y_d)^T \in \wedge_d$, and have a record of the positions of the dynamically evolving point vortices  $\b x_i \in \lbrace \wedge_0(t_n) \rbrace_{n=0}^{n_{t}}$, with know vorticity $\Gamma_{\b i}$. This is the reference solution and data. 
\item The drift components of velocity $\b u^{\delta}(\b x_{\b d},t_n)$ are estimated at weather stations by the Biot-Savart kernel \cref{eq: global biot savart} using known observed positions $\b x_{\b i}\in \wedge_0(t_n)$ of the point vortices at times $t_n$. 
\begin{align}
    \b u^{\delta}(\b x_{\b d}, t_n) = \sum_{\b i\in \wedge_0 } K_{\delta}(\b x_{\b d}- \b x_{\b i}(t_n))\omega_0(\b X_{\b i})h_x h_y, \quad \forall t_n \in \mathcal{T}, \quad \forall \b x_{\b d}\in \wedge_{d}.
    \end{align}

    \item 
    The difference $(\b u^{m}_{\b d} - \b u^{\delta}(\b x_{\b d}, t_n))  \Delta t$ between the measured velocity, and the drift reconstructed velocity at weather stations is taken $\forall t \in \mathcal{T}_n$. 
    Upon appropriate (invertible) vectorisation this is turned into a $n_t\times n_d$ matrix $M\in 
    \mathbb{R}^{n_t\times n_d}$
    where there are $n_d$ weather stations, and $n_t$ observation instances in time. Our specific vectorisation in space is the following operation,
    \begin{align*}
M_{n,:}  = [\operatorname{vec}( \operatorname{vec}(u^{m}_{\b d}(\b x_{\b d},t_n)),\operatorname{vec}(v^{m}_{\b d}(\b x_{\b d},t_n) ) ) - \operatorname{vec}( \operatorname{vec}(u^{\delta}(\b x_{\b d},t_n)),\operatorname{vec}(v^{\delta}(\b x_{\b d},t_n) ) ) ] \Delta t .
\end{align*}
 This $M$ is the data anomaly matrix, representing the effect of the stochastic velocity and driving signal on the weather stations.
\item  A common post-processing step in an SVD/PCA/EOF procedure is the removal of the row or column mean, such that the matrix has zero row sum or column sum. Since we are working with a $n_t\times n_d$ matrix, we remove the column(time mean) $TM(M):= \b e_{n_{t}} M / n_t$, from the data matrix in the following manner $M' = M - \b e_{n_{t}} \otimes (\b e_{n_{t}} M /n_t) $. The time mean difference observed from data will be denoted by $\bar{\b v}^{r}$.

\item One performs the truncated SVD(\cite{halko2011finding}) of the time mean removed data anomaly matrix $$M' = U_t\Sigma_t V_t^T=(c^{-1}U^c_t) (c\Sigma_t (V^c_t)^T).$$ We have re-scaled the construction by a constant $c\in \mathbb{R}$, such that the $k$-th column of $U = c^{-1}U^c_t$ has variance aligning with the timestep between observations, and the removal of time mean has normalised the data. For incremental data  $\operatorname{var}(U) = \Delta t$, and $\mathbb{E}
(U) = 0$.

Here the $p$-th row of $V_t^T\in \mathbb{R}^{n_t\times n_d}$, forms the $p$-th vectorised spatial eigenvectors $ \tilde{\b \xi}_p(\b x_d)$ effect on weather station data $\wedge_{\b d}$. The $(p,p)$-th element of $\Sigma_t$ denotes the corresponding singular value $\sigma_p$. The matrix product $U_t = M'V^{T}_t\Sigma^{-1}_t$, gives $U$ whose $p$-th column is the $p$th recovered path $\Delta W^{r}_{p}$ over the time window $\mathcal{T}$. For information about the SVD see \cref{remark:SVD:notes}.

\item Output: $\lbrace \sigma_{p}\tilde{\b \xi}_{p} \rbrace_{p\in [P]}$, $\bar{\b v}^r$, $\lbrace W^{r}_{p}\rbrace_{p\in [P]}$, are recovered from $\Sigma V^T,\b e_{n_{t}}M/n_{t},U$, respectively. 

\end{enumerate}
\end{method}

In practice, reconstruction is performed as to transform the discrete set of points $\b x_{\b d}\in \wedge^d$ and their reconstructed evaluation $\sigma_p \tilde{\b \xi}_p(\b x_{\b d})$, $\bar{\b v}^r(\b x_{\b d})$, at $\forall \b x_{\b d}\in \wedge^d$ into continuous fields $\sigma_p  \tilde{\b \xi}_p(x,y), \bar{\b v}^r(x,y)$. This interpolation step is required for the evolution of unstructured points in the calibrated inviscid vortex method. We use Fourier interpolation in the understanding, that specific to this work we expect periodic smooth basis functions and assume the data remains within the weather station grid, see \cref{remark:fourier interpolation:notes} for specifics in such an interpolation procedure. We propose two ensemble forward methods based on \cref{method:TSDVWSD}, one including the time mean drift (\cref{method:with}), and one without (\cref{method:without}).

One does not always have access to Eulerian weather-station data. It may be desirable to estimate parameters only from ``tracer" positional values such as evolving buoys or inherent state values in the evolving forward model. Furthermore, the values at weather stations need to be constructed into continuous fields, which require evaluation by interpolation in the forwards ensemble model during runtime, the cost of such a procedure scales badly (squared) with the number of weather-stations used see \cref{remark:cost of interpolation:notes}.
 
 It may be advantageous to avoid this computational runtime problem (particularly for large number of stochastic basis functions) or lack of weather-station data by setting the problem as a minimisation problem with a predefined basis. This will turn interpolation into evaluation in the forward model, resulting in a much faster ensemble method. However, the calibration problem is phrased as a significantly more costly nonlinear optimisation problem, described below (in the specific context of an inviscid vortex method).




\begin{method}[B(SPDEE)wrtFM]\label{method:backprop ensemble learning stage} Backpropagation through SPDE Ensemble with respect to diffusion parameters as to minimise a forecasting metric.
\begin{enumerate}
\item Data Collection; Record the positions of the dynamically evolving point vortices as data $\lbrace \b x^{*}_{\b i}\in \wedge_{0}(t_n)\rbrace_{n=0}^{n_t}$. Here the astrix superscript denotes data. This can be stored as a $n_t \times 2n_v$ matrix (positions in 2d have two components). 
\item Generate an entire ensemble run, over the time window of interest, recording all state variables aligning with observation instances in time, using $E_o$ number of proposed Ensemble members over an observation window $\mathcal{T}_o$. This is defined as a vectorised ensemble function denoted $F_E$, defined by going forwards in time with a forward discrete model of the following type, 
\begin{align}
&\b d_{t} \b x_{\b i}(t) = \b u^{\delta}(\b x_{\b i}(t),t) dt + \sum_{p=1}^{P}\theta_p \b \xi_{p}(\b x_{\b i}(t))\circ dW_e^{p}(t), \quad \forall \b i \in  \wedge_{0}, \forall t_n \in \mathcal{T}_o, \forall e \in [E_o].
\end{align}
This is done for all time steps in an observation window $\forall t_n \in \mathcal{T}_{o}$, and for $E_o$ realisations of $P$-dimensional Brownian motion from the initial condition. The output of the vectorised ensemble function $F_E$ is a $2 n_v \times n_{\mathcal{T}_{o}} \times E_o$ matrix, generated by the input of a $n_{\mathcal{T}_{o}} \times E_{o} \times P$ sized Gaussian random variable ``fed" as a component into a stochastic ensemble forecast model. This can be described heuristically as follows
\begin{align}
    F_{E}: \mathbb{R}^{P} \times \mathbb{R}^{n_{\mathcal{T}_{o}}\times E_{o} \times P} \times\mathbb{R}^{2n_v}\times\mathbb{R} \times \mathbb{R} \times \mathfrak{X}(\mathbb{R}^2)^P \times ... \mapsto \mathbb{R}^{2 n_v \times n_{\mathcal{T}_{o}} \times E_{o}}\\
F_{E}(\lbrace \theta_{p}\rbrace_{p=1}^{P}; \lbrace \Delta W^{p}_{e,n}\rbrace_{e,p,n\in [E_{o}],[P],[n_t]}; \lbrace\b x_{\b i}\rbrace_{\b i \in \wedge_0(t_0)};\delta,h,\lbrace \b \xi_{p}\rbrace_{p\in [P]}, ... ) = \text{ensemble forecast},
\end{align}
Where we have suppressed additional inputs in this function, as all but the first component (and perhaps the second) does not improve clarity. 
\item We then define the following observation averaged continuous rank probability score loss function, taking in the space-time observations and the ensemble forecast forward model
\begin{align}
L:\mathbb{R}^{2 n_v \times n_t }\times \mathbb{R}^{2 n_v \times n_t \times E_{o}} \mapsto \mathbb{R},  \quad L:= \frac{1}{n_t 2 n_v}\sum_{\forall n\in \mathcal{T}}\sum_{\forall \b x_{\b i}(t_n) \in \wedge_0(t_n)} \hat{\operatorname{crps}}(\b x_{\b i,t_{n}}^*;(F_{E})_{\b i,t_n}) \label{eq:crps_loss}
\end{align}
Where $\hat{\operatorname{crps}}:\mathbb{R}^{2 n_v \times n_t }\times \mathbb{R}^{2 n_v \times n_t \times E_{o}} \mapsto \mathbb{R}^{2 n_v \times n_t}$ denotes a vectorised continuous rank probability score estimator approximating the regular CRPS value over the space and time observations. The notation $(F_E)_{\b i,t_{n}}$ is used to indicate we compare the $E_o$ sized ensemble forecast at the position $\b x_{\b i}$, at time $t_n$, to the data point $\b x^*_{\b i,t_{n}}$ at the same location and temporal instance. For a single observation in space-time $x\in \mathbb{R},z\in \mathbb{R}^{E_{o}}$ this is done using the following formula
\begin{align}
\hat{\operatorname{crps}}(y, z) = \frac{1}{E_{o}}\sum_{e\in [E_{o}]}|y - z_e| - \frac{1}{E_{o}(E_{o}-1)}\sum_{(i,j)\in [E_{o}]\times [E_{o}]}|z_i-z_j|
\end{align}
see \cite{zamo2018estimation}, \cite{gneiting2011comparing}, and \cite{ferro2014fair}, for more insights into this estimator and its relationships to other estimators of the CRPS. See \cref{remark:crps:notes} for further insights and more detailed references as to the importance of the CRPS score in forecast verification. 
\item It is assumed that the discrete ensemble forecast model $F_E$ and the loss function is differentiable with respect to the parameters $\lbrace \theta_p \rbrace$, so one can compute the gradient, and perform (nonlinear) optimisation (e.g. gradient descent) to minimise the Loss(CRPS estimate using $\mathcal{T}_o$, $E_o$), through back-propagation. Should this converge, this is a methodology to minimise the CRPS average of an ensemble forecast, it is open to whether this can recover parameters such as $\lbrace\theta\rbrace_{p\in [P]}$ due to non-uniqueness.
\end{enumerate}
\end{method} 
\begin{remark} When $E_o=1$ \cref{method:backprop ensemble learning stage} is equivalent to minimising the mean absolute error, between a proposed solution path and the data. It is possible to interpret the above method as a stochastic version of an ensemble 4DVAR with a CRPS estimator loss. 
\end{remark}




\subsection{Methodology justification}
Mathematical motivation for the generation of the DAM (in \cref{method:TSDVWSD}), and adding the time mean back in (\cref{method:with}), can be justified by considering \cref{method:TSDVWSD} applied in the context of a twin experiment described below.

\begin{enumerate}
    \item In the context of a twin experiment where the data is generated by observing a stochastic model with known parameters $\lbrace \theta_{p}^{\ast} \rbrace_{p\in[P]}$, using a normally distributed driving signal $W^{\ast}$ assumed free of measurement error. The recorded total velocity field $\b u_{d}$ that would be seen by an observer at a weather station $\b x_{d}$ at time $t_n$ is assumed to be measurable in the following form 
    \begin{align}
            \b u_{\b d}(\b x_{d},t_{n}) = \b u^{\delta}(\b x_{\b d};\lbrace\b x_{\b i}\rbrace_{\b i \in \wedge^{\b i}_0(t_n)}) + \sum_{p=1}^{P}\theta_p^{*}\b \xi_{p}(\b x_{\b d})\Delta W_n^{\ast}/\Delta t + \b D(\b x_{\b d}),\quad \forall \bx_{d}, \quad \forall t_{n}\in \mathcal{T}_n. \label{eq: observed velocity in twin}
    \end{align}
    Where the Biot-Savart kernel \cref{eq: global biot savart} is used for the computation drift component of velocity $\b u^{\delta}(\b x_{\b d})$ induced by the point vortices at $\b x_{\b i }\in \wedge_0(t_n)$, and direct evaluation is assumed on $\b \xi_p$. Here we have divided the stochastic contribution to the velocity by $\Delta t$ as to represent how such a wind field would be measured in practice. We have also included the presence of an additional (time-independent) drift term $\b D$. 
    \item \Cref{method:TSDVWSD}, forms rows of the DAM from $\Delta t (\b u_{\b d}(\b x_{\b d},t_{n}) - \b u^{\delta}(\b x_{\b d};\lbrace\b x_{\b i}\rbrace_{\b i \in \wedge^{\b i}_0(t_n)}) $, which in the context of a twin experiment represents the discrete effect of $\b D\Delta t+\sum_{p=1}^{P}\theta_p^{*}\b \xi_{p}(\b x_{\b d})\Delta W_n^{\ast}$ for all measurement times (see \cref{eq: observed velocity in twin}). 
    \item  Let $ \Delta \b W^{*}_n \in \mathbb{R}^{P\times n_t}$, be a $(P,n_t)$-matrix made up of the $P$ dimensional sampled Brownian motion over $\lbrace t_{n} \rbrace_{n=1}^{n_t}$ used to generate the data. Let $\Xi \in \mathbb{R}^{d\times P}$, be a matrix whose $p$-th column is defined by the vertically stacked components of (vectorised) stochastic velocity contribution evaluated at the $d = n_c \times n_c$ weather stations of interest. Let $\hat{\b D}\in \mathbb{R}^{d}$, denote the vectorised drift effect on particle positions. Let $\b e_{n_{t}}$ denote a vector of ones length $n_{t}$, and $\otimes$ the outerproduct. 
    Then $M = \Delta t \b e_{n_{t}}\otimes\hat{\b D} + (\Delta \b W^{\ast})^T \Xi^T  \in \mathbb{R}^{n_t\times d}$ is the DAM observed in the twin experiment whose $n$th column is the stochastic contribution of velocity at the weather stations.
    \item The SVD procedure in \cref{method:TSDVWSD} finds an alternative representation of the effect of \begin{align}
        \Delta t \b e_{n_{t}}\otimes\hat{\b D} + (\Delta \b W^*)^T \Xi^T = M = e_{n_t} \otimes (e_{n_t} M /n_t)+ U\Sigma V^T, \label{eq:decomposition of the dam}
        \end{align} interpretable through PCA as the reconstruction of an efficient basis to explain the covariance structure over the time window of interest.
    \item A Stratonovich-Taylor expansion of the stochastic particle trajectory map, when evaluated at the weather stations reveals 
    \begin{align}
        \b x^{n+1}_{\b d} &= \b x^{n+1}_{\b d} + \b u^{\delta}(\b x^{n}_{\b d},t^{n})\Delta t +\b D\Delta t + (\Delta \b W^T(t^n) \Xi^T)^T + H.O.T,
    \end{align}
    where $\Delta \b W(t^n)$ is the $n$-th collumn. The
    substitution of the alternative representation in the other experiment \cref{eq:decomposition of the dam} gives
    \begin{align}
        \b x^{n+1}_{\b d} &= \b x^{n+1}_{\b d} + \b u^{\delta}(\b x^{n}_{\b d},t^{n})\Delta t +  (\b e_{n_t} \otimes (\b e_{n_t} M /n_t)+ U_n \Sigma V^T )^T + H.O.T. 
    \end{align}
    Where $U_n$ denotes the $n$-th row of $U$.
    Upon appropriate time rescaling we observe justification for the addition of a time mean, seen in \cref{eq: time mean equation} and \cref{method:with}. 
\end{enumerate}

The time mean drift term from data $(\b e_{n_t} M /n_t)$ can be well motivated to represent drifts $\hat{\b D}$ not present in the underlying model. One could foresee application in compensating systematic measurement error in sensing devices, or correcting for an unrepresented Itô-Stratonovich correction.  The time mean drift term from data $(\b e_{n_t} M /n_t)$ could also be seen as a parameterisation technique for representing unresolved drift terms, arising from fast dynamics \cite{diamantakis2024evy} or unresolved physics. However, if the model has no $\b D=0$, one does not necessarily get $(\b e_{n_t} M /n_t)=0$. An additional nonphysical drift can be observed, associated with a statistical error from sampling from the data distribution. The inclusion of the observed time mean drift term may bias towards a specific realisation of the data distribution, and not necessarily improve forecast skill. 

One of the objectives of this paper will be to numerically test the potential benefit for using an observed time mean drift $(\b e_{n_t} M /n_t)=0$ attained from \cref{method:TSDVWSD}. In the context of data arising from no time mean drift $D=0$. In the context where $D$ represents an Itô-Stratonovich sized drift term. In the context of data with a drift $D$ representing the effect of physical processes. These ideas will be tested using datasets(1,2), datasets 3 and datasets 4 respectively, introduced later. We use a twin experiment in which the aim is to re-simulate the training data, as well as computing forecast verification metrics on hidden testing data to account for statistical error associated with sampling the data distribution.


Mathematical motivation for \cref{method:backprop ensemble learning stage} and \cref{Method:ensemble_learning} is fairly transparent. The Continuous Ranked Probability Score (CRPS) is an example of a probabilistic forecast metric commonly used to assess ensemble forecast skill, a lower CRPS score indicates better forecast skill. The CRPS is probabilistic and compares the cumulative distribution function of the forecast with the observation values. In the context of an ensemble forward model, a lower CRPS score serves as an indicator of enhanced forecasting skill. In practice, this requires estimation over many observations, as taken into account with the loss in \cref{method:backprop ensemble learning stage}. For additional detail motivation and references regarding the CRPS see \cref{remark:crps:notes}.

\subsection{Ensemble methods}\label{sec: ensemble methods}

This section contains a list of ensemble methods that will be tested, these are forward models, some requiring estimation of $\lbrace\theta_p \rbrace_{\forall p}$, some require generation of a basis $\lbrace\tilde{\b \xi_p}\rbrace_{\forall p}$. Methodology for such estimation has been described in the previous \cref{sec:svd and ml  Calibration methodology}. 

\begin{ensemble method}[Persistence]\label{method: persistence} We predict an ensemble whose particles remain at their initial conditions $\wedge_0(t) = \wedge_0$ for the entire time interval of interest. 
\end{ensemble method}

\begin{ensemble method}[RIC (Random initial condition perturbation)]\label{method:Random} We initially perturb each particle position by scaled samples from the normal distribution, then we run forwards with the deterministic model to generate an ensemble. (A sensible magnitude (giving small CRPS) perturbation was searched for through trial and error and found to be of the type $0.001 N(0,\mathbb{I})$.) 
\end{ensemble method}

\begin{ensemble method}[Perfect model]\label{method: perfect ensemble}
We propose the SPDE used to generate the synthetic data, as a forecast model. This involves knowing true parameters $\lbrace\theta^* \rbrace_{p\in [P]}$ and running an ensemble forecast with new samples from the normal distribution. 
\end{ensemble method}



\begin{ensemble method}[TSVDWD without time mean] \label{method:without} We perform \cref{method:TSDVWSD} to obtain a basis for noise $\lbrace \sigma_{p}\tilde{\b \xi}_{p}(\b x)\rbrace_{\forall p}$. We then use the SSP33 stochastic integrator to run the regularised integrodifferential model for particle trajectories
\begin{align}
&\b d_{t} \b x_{\b i}(t) = \b u^{\delta}(\b x_{\b i}(t),t) dt + \sum_{p=1}^{P}\sigma_p \tilde{\b \xi_{p}}(\b x_{\b i}(t))\circ dW^{p}(t), \quad \forall \b i \in  \wedge_{0},
\end{align}
Where $\b u^{\delta}(\b x_{\b i}(t),t)$, is computed as before using \cref{eq: regularised biot savart}, and Fourier interpolation (\cref{remark:fourier interpolation:notes}) is used to evaluate $\tilde{\b\xi}$, at points. 
\end{ensemble method}

\begin{ensemble method}[TSVDWD with time mean] \label{method:with}
 We perform \cref{method:TSDVWSD} to obtain a basis for noise $\lbrace \sigma_{p}\tilde{\b \xi}_{p}(\b x)\rbrace_{\forall p}$, and a time mean effect $\bar{\b v}(\b x)$. We then use the same SSP33 stochastic integrator to run the regularised integrodifferential model for particle trajectories
\begin{align}
&\b d_{t} \b x_{\b i}(t) = \b u^{\delta}(\b x_{\b i}(t),t) dt+\bar{\b v}(\b x_{\b i}(t)) dt + \sum_{p=1}^{P}\sigma_p \tilde{\b \xi_{p}}(\b x_{\b i}(t))\circ dW^{p}(t), \quad \forall \b i \in  \wedge_{0}, \label{eq: time mean equation}
\end{align}
where $\b u^{\delta}(\b x_{\b i}(t),t)$, is computed as before using \cref{eq: regularised biot savart}, and the Fourier interpolation described in \cref{remark:fourier interpolation:notes} is used is used in the evolution of the points by the additional deterministic drift term $\bar{\b v}$ and stochastic terms $\lbrace\tilde{\b\xi}\rbrace_{ p\in [P]}$. 

\end{ensemble method}


\begin{ensemble method}[Backpropagation ensemble approach]\label{Method:ensemble_learning}
We take the parameters $\lbrace \theta_p\rbrace_{p\in [P]}$ that minimise the CRPS mean loss after \cref{method:backprop ensemble learning stage} is performed over $\mathcal{T}_o$ with $E_{o}$ ensemble members, and run the trained ensemble method $F_E$ forward 
\begin{align}
&\b d_{t} \b x_{\b i}(t) = \b u^{\delta}(\b x_{\b i}(t),t) dt + \sum_{p=1}^{P}\theta_p \b \xi_{p}(\b x_{\b i}(t))\circ dW_e^{p}(t), \quad \forall \b i \in  \wedge_{0}, \forall t_n \in \mathcal{T}, \forall e \in [E].
\end{align}
Where $\b u^{\delta}(\b x_{\b i}(t),t)$, is computed as before using \cref{eq: regularised biot savart}, and the vectorfields $\lbrace \b \xi_p \rbrace_{p\in[P]}$ are directly evaluated rather than Fourier interpolated. This could potentially be interpreted as a stochastic ensemble version of 4DVAR with a CRPS loss. 
\end{ensemble method}


\section{Numerical experiments.} \label{sec:experiments}

\subsection{Twin experiment frameworks}
In operational practice (in, say, weather prediction), the state values such as temperature come from an unknown distribution and are recorded using measurement devices with the addition of measurement noise. However, to assess the proposed data assimilation methodology a known reference dataset should be predefined beforehand. This naturally leads to the concept of a twin experiment framework, a common practice in both weather forecasting and inverse modelling communities. We will now (in the next paragraph) describe how by fixing the driving path, we can perform a twin experiment for the SVD calibration of a stochastic fluid system. We then, in the proceeding paragraph describe another method of validation, in which the underlying distribution of the observation signal is assumed known. This type of testing can alleviate errors associated with sampling data from the unknown distribution. 


A reference trajectory is assumed known, and computed by fixing all parameters $\lbrace\theta^{*}_p\rbrace_{p=0}^{p=P}$, $\lbrace\lbrace \Delta W_p(t_n)\rbrace_{p=0}^{p=P}\rbrace_{n=0}^{N}$ and running the stochastic forward model over a time window $t\in[0,T]$. Synthetic measurements (at weather-stations) are then collected by sampling values from this reference trajectory. Finally, the data assimilation technique of interest \cref{method:TSDVWSD} is implemented as to attain $\lbrace \sigma^{r}_{p}\rbrace_{p=0}^{p=P}$, $\lbrace\lbrace \Delta \tilde{W}_p(t_n)\rbrace_{p=0}^{p=P}\rbrace_{n=0}^{N}$, and $\lbrace \bar{\b v}^{r} \rbrace$. Using these ``recovered" parameters, we generate a new output trajectory, for the evolution of points. We then can compare the output trajectory to the reference trajectory. This allows the SVD calibration accuracy to be assessed. These tests will be performed and assessed using datasets 1,3,4 and 2, using $P=\lbrace 1,5\rbrace$ respectively.
In this setting the twin experiment is not viewed in the context of verifying the calibration of a stochastic model, but as an assessment of the method viewed as a data assimilation procedure in which the reference ``training" trajectory is aimed at being captured as accurately as possible. 

Going further than this, suppose for testing/validation purposes that we know more than just a single reference trajectory, but the entire reference distribution. Namely, we know the stochastic forward model used to generate data and its parameters (not necessarily the one proposed for modelling). In this setting, we can account for the additional sampling error associated with drawing data from the underlying distribution. To do so in practice we generate 1000 realisations of the stochastic forward model used for data. These 1000 realisations of the stochastic forward model will be treated as hidden testing datasets for which ensemble forecast verification metrics can be employed. In this scenario, CRPS scores arise for data for which the model has not been trained, this can help distinguish stochastic sample path model error associated with sampling the data distribution. This can be thought of as the verification/validation of unseen/hidden test data.

\subsection{Setup: Generation of synthetic data}\label{sec:synthetic data generation}

This subsection contains specific details about the generation of four reference datasets we wish to calibrate from.  Dataset 1 is made with a realisation of a parametrised stochastic model with one basis function. Dataset 2 is made with a realisation of a parametrised stochastic model with 5 basis functions. Dataset 3 is made with a realisation of a stochastic model, but the model has a predefined additional drift, replicating model-data mismatch. Dataset 4 contains the same set-up as Dataset 3 but with a different predefined drift, larger in magnitude replicating more realistic model data mismatch. Per dataset, we also compute an additional 1000 corresponding realisations for testing purposes.

\begin{figure}[hbt!]
\centering
\begin{subfigure}[t]{0.32\textwidth}
\centering
\includegraphics[width=.95\textwidth]{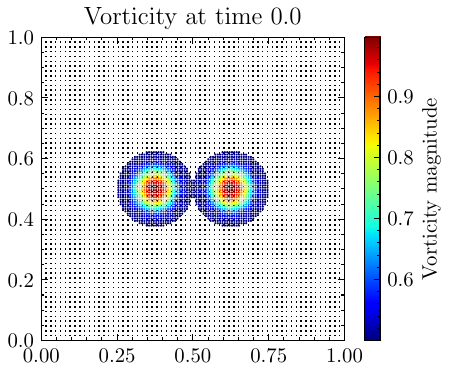}
\end{subfigure}
\begin{subfigure}[t]{0.32\textwidth}
\centering
\includegraphics[width=.95\textwidth]{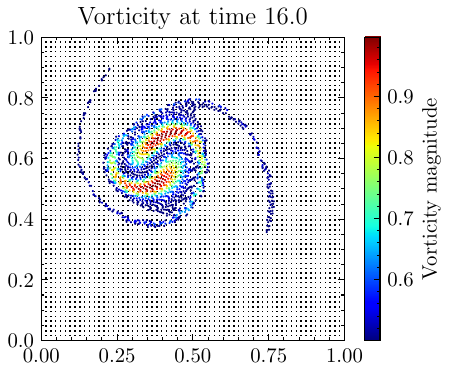}
\end{subfigure}
\begin{subfigure}[t]{0.32\textwidth}
\centering
\includegraphics[width=.95\textwidth]{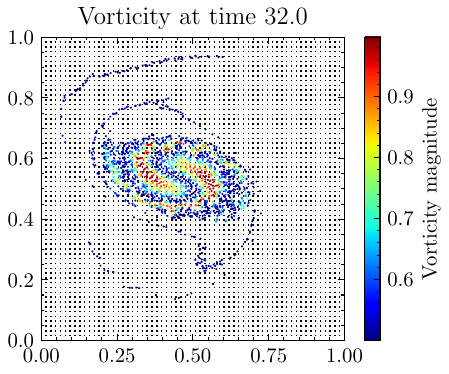}
\end{subfigure}
\begin{subfigure}[t]{0.32\textwidth}
\centering
\includegraphics[width=.95\textwidth]{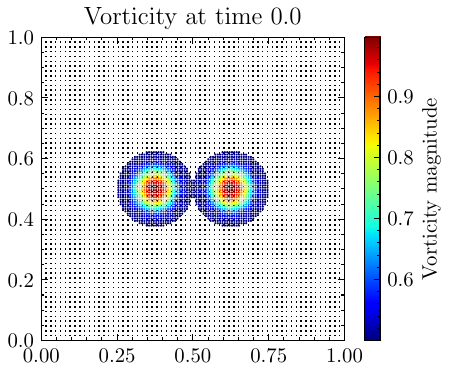}
\end{subfigure}
\begin{subfigure}[t]{0.32\textwidth}
\centering
\includegraphics[width=.95\textwidth]{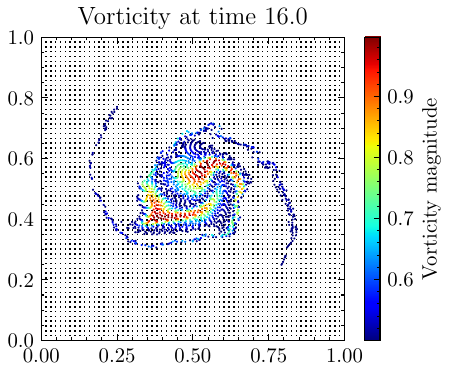}
\end{subfigure}
\begin{subfigure}[t]{0.32\textwidth}
\centering
\includegraphics[width=.95\textwidth]{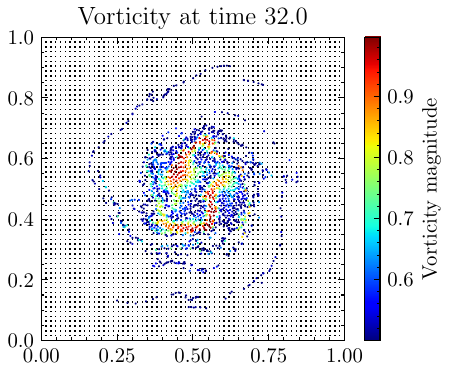}
\end{subfigure}
\begin{subfigure}[t]{0.32\textwidth}
\centering
\includegraphics[width=.95\textwidth]{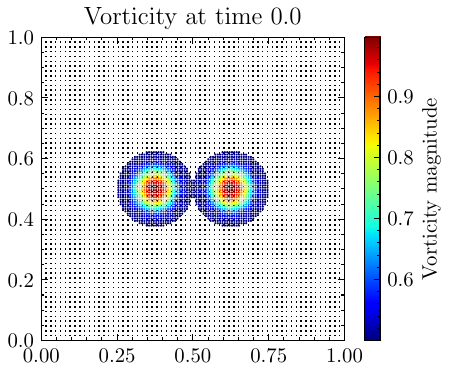}
\end{subfigure}
\begin{subfigure}[t]{0.32\textwidth}
\centering
\includegraphics[width=.95\textwidth]{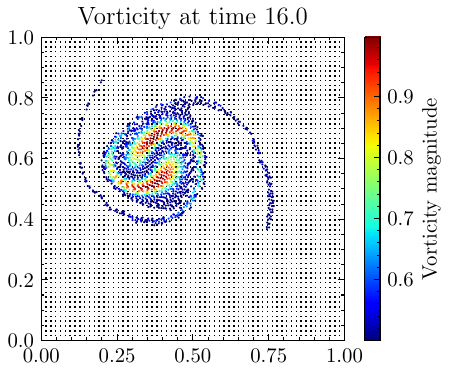}
\end{subfigure}
\begin{subfigure}[t]{0.32\textwidth}
\centering
\includegraphics[width=.95\textwidth]{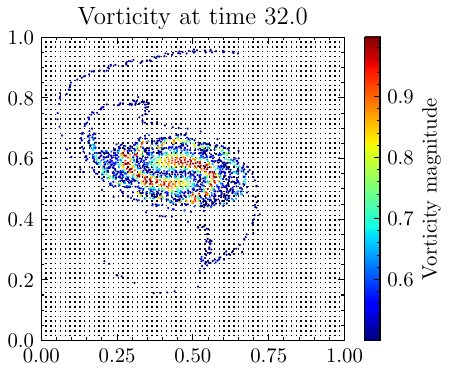}
\end{subfigure}
\begin{subfigure}[t]{0.32\textwidth}
\centering
\includegraphics[width=.95\textwidth]{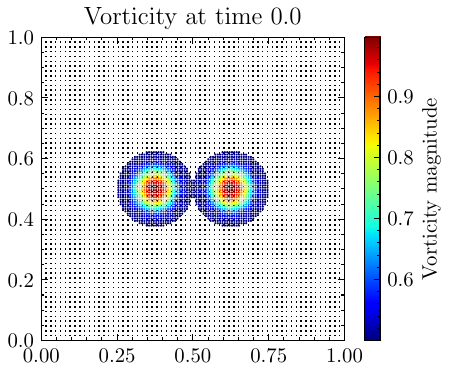}
\end{subfigure}
\begin{subfigure}[t]{0.32\textwidth}
\centering
\includegraphics[width=.95\textwidth]{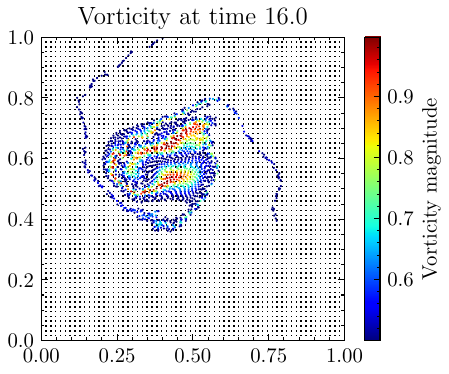}
\end{subfigure}
\begin{subfigure}[t]{0.32\textwidth}
\centering
\includegraphics[width=.95\textwidth]{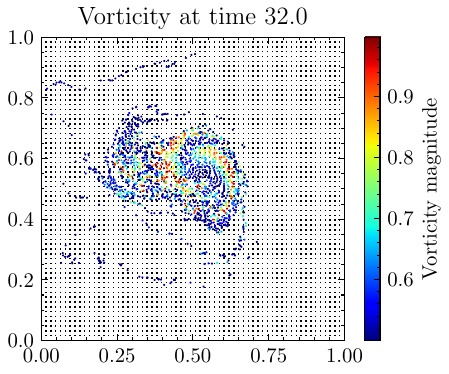}
\end{subfigure}
\caption{ Shown in each row is a scatter plot of point vortex positions whose vortex strength is indicated by the non-perceptually uniform colourmap jet, as to highlight finer structures in the flow. Black square dots denote weather stations from which data is collected. In row one we plot snapshots of the point vortex (at t = 0,16,32) for dataset 1 generated with one basis function. Row two corresponds to dataset 2 driven with five basis functions. Row three corresponds to dataset 3, generated with one basis function and an additional Stratonovich-Itô drift. Row four corresponds to dataset 4, generated with one basis function and a prescribed deterministic drift, representing additional unresolved processes. These datasets contain the effect of different prescribed physical processes (Stochastic diffusion and drift terms). This work aims to extract and parameterise the effects of these hidden processes, for use with a stochastic ensemble forward model.}\label{fig:total points evolving.}
\end{figure}

 The initial condition $\omega_{0}(x,y)$, is constructed from two compactly supported circular regions with radius $R=1/8$ of non zero vorticity in the following way   
\begin{align}
\omega_{0}(x,y) :=
\begin{cases}
1/2+1/2\left(1 - (\frac{r_1}{R})^2\right)^3, \quad r_1 = ((x- \frac{1}{2}-R)^2+(y-\frac{1}{2})^2)^{1/2} \leq R, \\
1/2+1/2\left(1 - (\frac{r_2}{R})^2\right)^3, \quad r_2 = ((x-\frac{1}{2}+R)^2+(y-\frac{1}{2})^2)^{1/2} \leq R, \\
0 \quad \text{else}.
\end{cases}
\end{align}
We use a initial $n \times m = 128\times128$ mesh over the domain $[0,1] \times [0, 1]$.
Resulting in a mesh spacing of $h = h_x = h_y = 0.0078125$
and regularisation parameter $\delta = 1.5 h^{3/4} = 0.03941702$.
We remove the point vortices with non-zero vorticity from the flattened $(x,y)$-meshgrid of points.
Resulting in $N_v = 1216$ points remaining from the $16384$ initially specified on the mesh. We use $n_t = 256$ timesteps on the time interval $t \in [0,32]$, with $\Delta t = 0.125$.  

In dataset 1, we consider idealised data generated from a parametrised run of the SPDE, with  a single $P = 1$ basis function $\theta_1 = 0.003$, given by 
\begin{align}
\theta_1 (\xi_{1}^{x},\xi_{1}^{y})^{T} = 0.003 (2 \pi \cos( 2 \pi y) , -2 \pi \cos(2 \pi x))^{T}. \label{eq: basis 1}
\end{align}
In dataset 2, we consider data generated from a parametrised run of the SPDE with 5 basis functions given by
\begin{align}
    \theta_p (\xi_{p}^{x},\xi_{p}^{y})^T = 0.0001  (2\pi \cos(2\pi p y), -2\pi \cos(2\pi p x))^T, \quad p\in [5]. \label{eq: basis 5}
\end{align}

In dataset 3, we create the data as a single realisation of the time mean included system described by \cref{eq: time mean equation}, where we prescribe the same basis function as dataset 1 \cref{eq: basis 1}, however we choose the following Stratonovich-Itô correction drift
\begin{align}
\bar{\b v} = - 0.003^2 4\pi^3(\sin(2\pi y) \cos(2 \pi x),  \sin(2 \pi x)  \cos(2 \pi y) )^{T}, \label{eq: ito stratonovich drift function}
\end{align}
in the underlying stochastic model that generates the data. To test the importance or non-importance of the Itô-Stratonovich correction in the generation of the data anomaly matrix.

In dataset 4, we create the data as a single realisation of the time mean included stochastic system with the same basis function \cref{eq: basis 1} to that of datasets 1 and 3 but use the following drift
\begin{align}
\bar{\b v} = 0.0003 (8 \pi \cos( 8 \pi y) , -8 \pi \cos(8 \pi x))^{T}, \label{eq: physical drift}
\end{align}
replicating some small-scale unresolved drift velocities not proposed in the stochastic forward model, but observed by the data.

We either use $\lbrace \Delta W_1(t_n)\rbrace_{n\in [n_t]}$ a $n_t$ sized sample from the normal distribution or $\lbrace \Delta W_p(t_n)\rbrace_{n\in [n_t], p\in [5]}$ a $(n_t\times 5)$ sized sample from the normal distribution for datasets (1,3,4) and 2 respectively. The resulting set of evolving points $\wedge_0(t)$, do not remain a Cartesian mesh after initial time. We also consider a $ 64 \times 64 $ Cartesian meshgrid $\wedge^{d}$, of fixed weather centers in a closed subdomain $[0,1] \times [0,1]$ of $\mathbb{R}^2$ for all time. Where the additional subscript $d$ denotes ``data", and indicates that this is a weather-station in which velocity data $\b u^{m}(\b x_{\b d}, t_n)$ is measured and recorded (see e.g. \cref{eq: observed velocity in twin}). 


Snapshots at $t=(0,16,32)$ of the stochastic solution dataset 1 is plotted in the first row of \cref{fig:total points evolving.}, generated with the addition of a single basis function \cref{eq: basis 1}. In the second row of \cref{fig:total points evolving.}, we plot snapshots of dataset 2, generated with five parametrised basis functions \cref{eq: basis 5}. In the third row of \cref{fig:total points evolving.} we plot snapshots of dataset 3, generated with a single basis function \cref{eq: basis 1} and a Stratonovich-Itô correction drift \cref{eq: ito stratonovich drift function}. In the fourth row of \cref{fig:total points evolving.} we plot snapshots of dataset 4, generated with a single basis function \cref{eq: basis 1} and a pre-prescribed drift function \cref{eq: physical drift} representing physical unresolved processes. Not plotted are an additional 1000 hidden testing/validation datasets per the above dataset.

\subsection{Results and discussion}

We apply the SVD approaches based on \cref{method:TSDVWSD}, to datasets 1,2,3,4 in a context of a twin experiment for the re-simulation of data. We apply the ensemble backpropagation method \cref{method:backprop ensemble learning stage} to only datasets 1,2, as we have not described the extension of the \cref{Method:ensemble_learning,method:backprop ensemble learning stage} to actively include explicit drift parameterisation. We then compute ensemble forecast verification metrics on the hidden test data, to see if the underlying distribution is well represented.

\subsubsection{Results: Dataset 1} \label{results for dataset one}

\paragraph{Twin experiment}
Synthetic dataset 1 $\lbrace \lbrace \b x_{\b i}(t_n)\rbrace_{\b i \in \wedge^{\b i}_{0}(t_n)} , \lbrace\b u^{m}(\b x_{\b d},t_n)\rbrace_{\b x_d\in \wedge_{d}} \rbrace_{n =0}^{n=n_t}$ described in \cref{sec:synthetic data generation} was generated using a single basis of noise (\cref{eq: basis 1}), whose snapshot (at $t=0,16,32$) of vortex positions is shown in the first row of \cref{fig:total points evolving.}.

\begin{figure}[!hbt]
\centering
\begin{subfigure}[t]{0.3\textwidth}
\centering
\includegraphics[width=.95\textwidth]{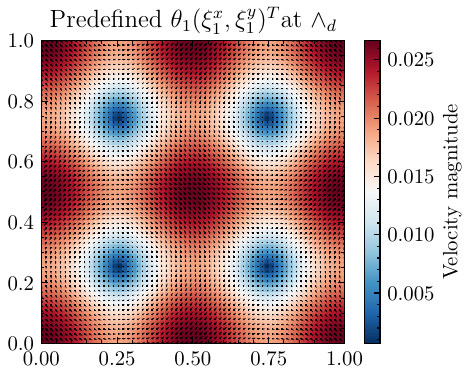}
\caption{\hfill}
\label{fig:predefined_theta_P1}
\end{subfigure}
\begin{subfigure}[t]{0.31\textwidth}
\centering
\includegraphics[width=.95\textwidth]{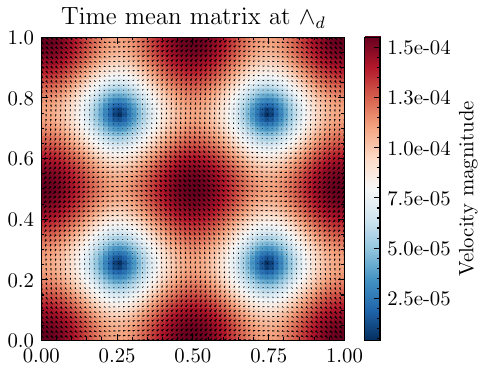}
\caption{\hfill}
\label{fig:recovered_timemean}
\end{subfigure}
\begin{subfigure}[t]{0.3\textwidth}
\centering
\includegraphics[width=.95\textwidth]{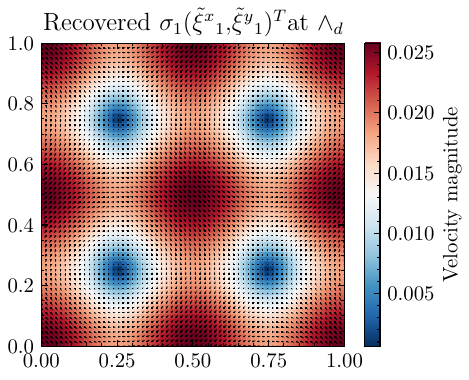}
\caption{\hfill}
\label{fig:recovered sigma p1}
\end{subfigure}
\begin{subfigure}[t]{0.3\textwidth}
\centering
\includegraphics[width=.95\textwidth]{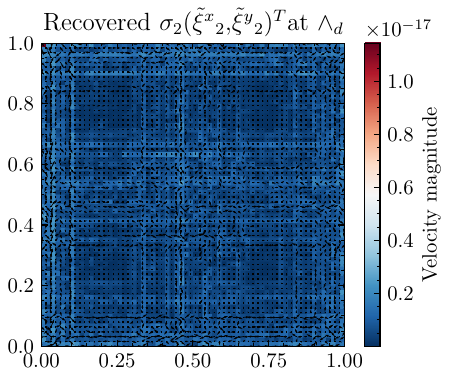}
\caption{\hfill}
\label{fig:recovered sigma p2}
\end{subfigure}
\begin{subfigure}[t]{0.35\textwidth}
\centering
\includegraphics[width=.95\textwidth]{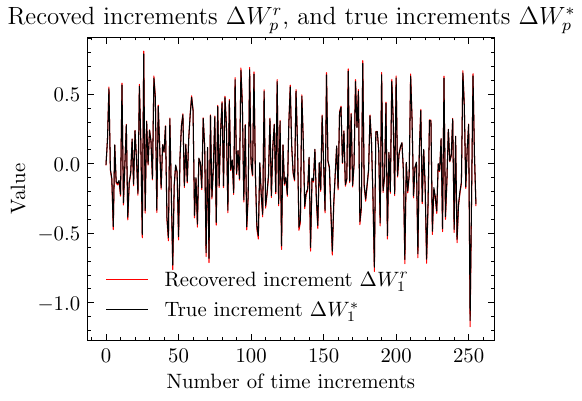}
\caption{\hfill}
\label{fig:recovered increments.}
\end{subfigure}
\caption{ Predefined basis of noise \cref{fig:predefined_theta_P1} (a), recovered time mean \cref{fig:recovered_timemean}(b), recovered first basis \cref{fig:recovered sigma p1}(c), second recovered basis \cref{fig:recovered sigma p2}(d), recovered first increment \cref{fig:recovered increments.}(e), after
method \cref{method:TSDVWSD} is applied on dataset 1. }
 \label{fig: recovery of one basis function}
\end{figure}

\Cref{fig: recovery of one basis function}(a) contains a plot of the basis vector-field $\theta_1 \b\xi_1(\b x_{\b d})$, used (in combination with the forward model) to generate synthetic dataset 1. Over 99.99 percent of covariance was explained by one basis function. \Cref{fig: recovery of one basis function}(b) contains the time average velocity field drift from the data. \Cref{fig: recovery of one basis function}(c) contains the recovered vector field $\sigma_1 \tilde{\b \xi}_{1}$ by \cref{method:TSDVWSD}. Visually \cref{fig: recovery of one basis function}(a) and \cref{fig: recovery of one basis function}(c) appear similar, and have agreement to $0.0467482$ in the relative $L^2$ norm. The next recovered basis $\sigma_2 \tilde{\b \xi}_{2}$ has machine precision magnitude plotted in \cref{fig: recovery of one basis function}(d). \Cref{fig: recovery of one basis function}(e) contains a plot of the recovered time increments $\Delta W_1^r$, and the driving signal increments $\Delta W_1^*$, $\Delta W_1^r$ and $\Delta W_1^*$ are nearly indistinguishable apart from a small difference in magnitude. They differ in the relative $L^2$-norm by 0.03818701. With the addition of the time mean it is possible to recover the data anomaly matrix to 2.00982e-13 using only one singular value, one singular vector and the recovered driving signal. Note the SVD decomposition could equivalently result in an opposite signed $\b \xi_1$, and opposite signed $\Delta W_1^r$ increments.

We have presented evidence indicating that the methodology in \cref{method:TSDVWSD} identified a Data anomaly matrix related to the effect of $\theta_1 \b\xi_1(\b x_{\b d})$ from the synthetic data.
    We speculate the small difference between $\sigma_1 \tilde{\b \xi}_{1}$ and $\theta_1 \b\xi_1(\b x_{\b d})$, is in part related to the removal of the time average, specific to the realisation of Brownian motion $\Delta W^*_1$ used in the synthetic data. This motivates the next test where we shall drive the solution with the recovered increments, recovered basis functions, and with or without the recovered time mean drift, all in comparison to dataset 1. We shall call this the re-simulation of data.

\begin{figure}[hbt!]
\centering
\begin{subfigure}[t]{0.325\textwidth}
\centering
\includegraphics[width=.95\textwidth]{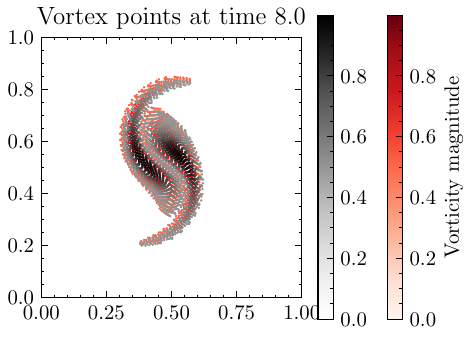}
\end{subfigure}
\begin{subfigure}[t]{0.325\textwidth}
\centering
\includegraphics[width=.95\textwidth]{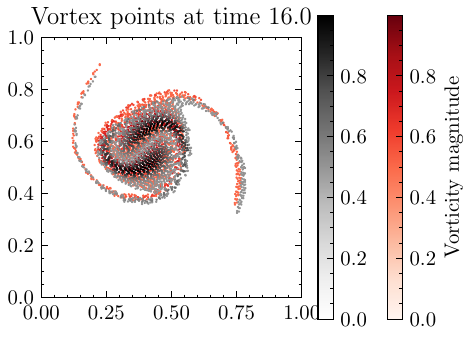}
\end{subfigure}
\begin{subfigure}[t]{0.325\textwidth}
\centering
\includegraphics[width=.95\textwidth]{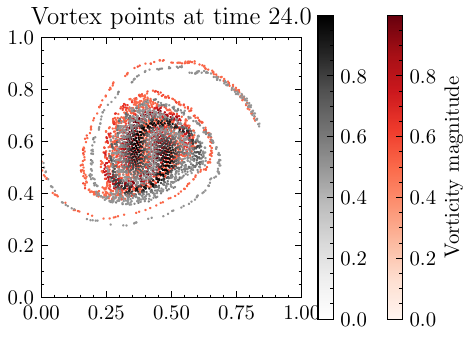}
\end{subfigure}\\
\begin{subfigure}[t]{0.325\textwidth}
\centering
\includegraphics[width=.95\textwidth]{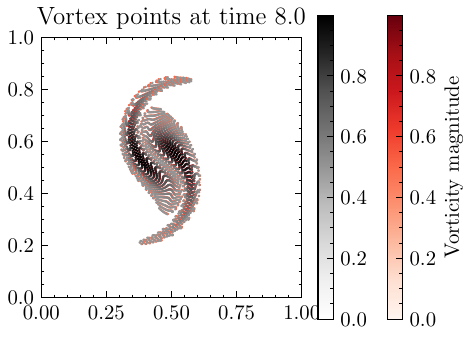}
\end{subfigure}
\begin{subfigure}[t]{0.325\textwidth}
\centering
\includegraphics[width=.95\textwidth]{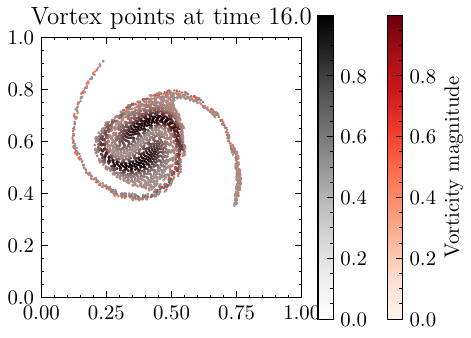}
\end{subfigure}
\begin{subfigure}[t]{0.325\textwidth}
\centering
\includegraphics[width=.95\textwidth]{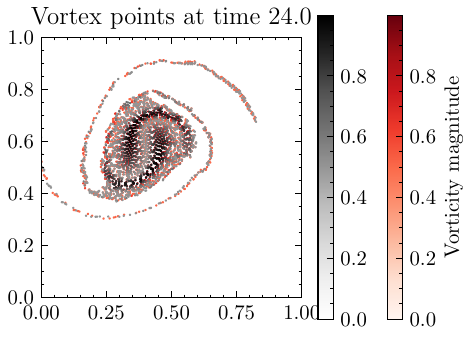}
\end{subfigure}
\caption{Dataset 1 $(P=1)$, reconstruction of data. Grey data points are the data. The red points are the recovered increment driven system. In the first row, the red points are evolved without using the additional time mean contribution using \cref{method:without}. The second row contains the same experiment but with the additional time mean drift using \cref{method:with}. We observe that the time mean velocity drift has notable effect, in the context of this re-simulation of data twin experiment. } \label{fig:dataset one driven solution. }
\end{figure}

\Cref{fig:dataset one driven solution. } contains the solution of the SPDE when driven by the recovered signal $\Delta W^r_1$ with recovered basis function $\sigma_1 \tilde{\b \xi_1}$ as compared with the original data using $\Delta W^*_1$ and $\theta^*_1 \b \xi^*_1$. In \cref{fig:dataset one driven solution. } the first 3 images in row one do not have the addition of the time mean, where as row two contains the time mean drift velocity. The relative (L2-spacetime) error of the reproduced time mean included solution is $0.01265$ the relative error of the recovered time mean not included solution is $0.05499$. 
As seen in \cref{fig:dataset one driven solution. } and measured by relative error of the reproduced solution the inclusion of the time mean was found to be significantly helpful in the re-simulation of the synthetic data in the context of a twin experiment.  The remaining non-zero difference in relative error norm $0.01265$ could be speculated to be caused by many things, such as small inaccuracies in Fourier interpolation growing over the course of the simulation run.



\paragraph{Verification of learning during training}
We train the ensemble back-propagation approach \cref{method:backprop ensemble learning stage} using a $E_{o}\in \lbrace 1,5 \rbrace$ sized member ensemble forecast over a smaller time window $\mathcal{T}_{o} = [8]/8$ such that the forward model creates a $E_{o}$ sized ensemble in which 8 SSP33 Runge Kutta steps are taken for each ensemble member. The CRPS-loss of the $E_o$ sized ensemble forecast over the time window $\mathcal{T}_o$ is denoted $\operatorname{crps}_{o}$. The parameters in the gradient descent algorithm used, are learning rate 1e-7, acceleration = true, max iterations 500, tolerance 1e-15, implemented using the ``jaxopt.GradientDescent" algorithm. We take the initial guess of parameters to be $\theta^g_1 = 10^{-12}\b e_p$, such that the initial CRPS score before training is essentially a measure of the forecast skill of an ensemble with $E_o$ deterministic models. 

\begin{figure}[hbt!]
\centering
\begin{subfigure}{0.45\textwidth}
\centering
\includegraphics[width=.9\textwidth]{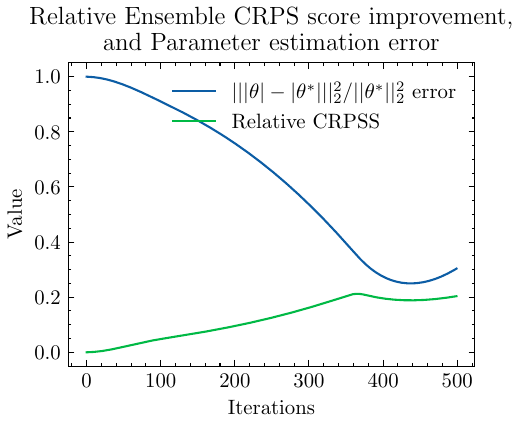}
\end{subfigure}
\begin{subfigure}{0.45\textwidth}
\includegraphics[width=.9\textwidth]{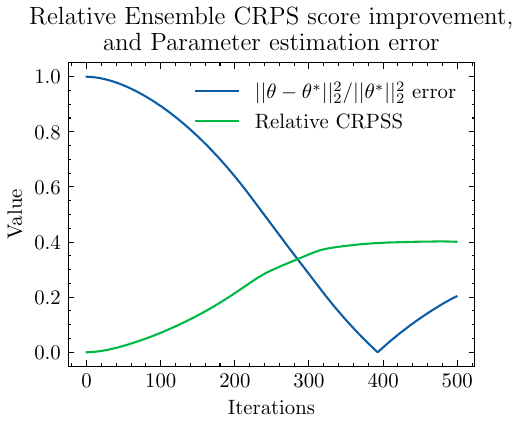}
\end{subfigure}
\caption{The green line is the relative CRPSS improvement as compared to the initial deterministic proposed ensemble. The blue line is the relative error associated with the magnitude in parameter estimation on training data. 
The figures are plotted for $E_{o}=1$, $E_o=5$, respectively. Both figures show a 30 to 40 percent increase in forecast skill over the deterministic proposed ensemble. Learned parameters are closer in magnitude to the true parameters used for the generation of synthetic data. However, there are situations in which the true parameter does not necessarily give a better CRPSS. }
\label{fig:losses}
\end{figure}

In \cref{fig:losses} we plot the Relative improvement in CRPSS$_{o}$ over the initial Deterministic ensemble forecast (30-40 percent improvement), and the relative error in parameter estimation magnitude, for $E_{o} \in \lbrace1,5\rbrace$ respectively. For $E_{o} \in \lbrace 1,5 \rbrace$, the true parameter value was $\theta^*_1 \in \lbrace 0.003,0.003\rbrace$, the initial guess is $\theta^g_1 = 10^{-12}$, after $500$ iterations we have learned parameter value
$\theta^r_1 \in \lbrace -0.00208943,0.00360890\rbrace$ respectively. The initial CRPS estimate is $\operatorname{crps}_o = \lbrace 0.0101590, 0.0101590\rbrace$, the after training the final CRPS estimate over the time interval $\mathcal{T}_o$ is $\operatorname{crps}_o \in  \lbrace 0.00809122,0.00608491\rbrace$ respectively, this CRPS is computed on a subset of the training data, $\mathcal{T}_o$. 


\paragraph{Forecast Verification for underlying distribution}
For dataset 1, an additional 1000 hidden validation/testing datasets were created, each testing dataset is the exact stochastic model used for the production of dataset 1, but run forwards with new normally distributed sampled increments. We then compute a 30-member ensemble forecast using a stochastic forward model trained/calibrated on dataset 1. 
Per hidden dataset, we estimate the CRPS over all space-time observations (using \cref{eq:crps_loss}) representing the likelihood of observing the test dataset from the ensemble forecast. We then take the mean CRPS over the entire 1000 test datasets. A lower mean CRPS value informally indicates a better likelihood that the hidden testing dataset comes from the trained ensemble methods 30-member ensemble forecast prediction. 

The raw averaged CRPS scores are displayed in the first column of \cref{table:total}, and the percent relative improvements in average CRPSS (calculable as $100(1-\text{CRPS(model 1)}/\text{CRPS(model 2))}$) are displayed in \cref{table: Relative skill hidden 1}, we observe
the following. The Without-time-mean ensemble (calibrated with method 3.1) outperformed the perfect
model by 0.04 percent. The Perfect model outperformed the With-time-mean model by 2.039 percent. The With-time-mean
model outperformed randomised initial conditions by 8.071 percent. The Randomised initial condition outperformed
persistence by 49.44 percent.

We conclude that the underlying distribution is best represented by the Perfect model and the Without-time-mean model performed similar in CRPS score. The With-time-mean model performed marginally worse, we speculate that this is because the observed drift was a sampling error rather than a systematic modelling error. Correcting for a sampling error, biased the solution towards the reference training data as seen in \cref{fig:dataset one driven solution. }, but was not helpful for representing (on average) the hidden 1000 testing datasets i.e. the underlying distribution.

\begin{table}[htp!]
\centering
\begin{tabular}{||c|c|c|c|c|c||}
\hline
scheme & RIC & With-time-mean & Without-time-mean & Perfect & Persistence \\
\hline
\hline
RIC & 0 & -8.779 & -11.09 & -11.04 & 49.44 \\
\hline
With-time-mean & 8.071 & 0 & -2.123 & -2.081 & 53.52 \\
\hline
Without-time-mean & 9.982 & 2.079 & 0 & 0.0413 & 54.48 \\
\hline
Perfect & 9.945 & 2.039 & -0.04132 & 0 & 54.47 \\
\hline
Persistence & -97.77 & -115.1 & -119.7 & -119.6 & 0 \\
\hline
\end{tabular}
\caption{Hidden dataset 1, ``Percent improvement" table of row scheme over the column scheme for representing the distribution, i.e. hidden datasets. Readable as follows, the With-time-mean ensemble is 8.071 percent better at representing the underlying distribution than the RIC ensemble. In practice this is the relative skill, hidden dataset averaged, observation averaged CRPS estimate, of the row scheme against the column scheme multiplied by a factor of 100.}
\label{table: Relative skill hidden 1}
\end{table}

In summary, \cref{table:total,table: Relative skill hidden 1} indicate that using the time mean drift is not helpful in representing the underlying distribution when the underlying distribution (e.g. hidden SPDE/SDE model) does not have an explicit time mean drift. This is fairly transparent as the time mean drift occurs only as a result of sampling data from the underlying distribution, and adding in the small observed drift biases towards the training dataset rather than compensating for a model-data drift mismatch. However, in practice one may not be able to tell the difference between model error and statistical sampling error, as one normally cannot resample from the underlying distribution. Biasing the model towards matching observed data with a time mean drift could be seen as a valid modelling assumption to make, and only decreased relative CRPSS by about 2 percent for this case. It should be noted that a sensible random perturbation of the initial condition performed remarkably well (10 percent worse CRPSS) at representing SALT-type data. Both \cref{method:with} and \cref{method:without} (with and without the time mean) outperformed both RIC and Persistence, and approached the CRPS score of the perfect model.

\subsubsection{Results: Dataset 2}
\paragraph{Twin experiment} For synthetic dataset 2 described in \cref{sec:synthetic data generation} we used $P=5$ velocity fields as a basis of noise \cref{eq: basis 5} for the generation of data, whose snapshot (at $t=0,16,32$) of vortex blobs is shown in the second row of \cref{fig:total points evolving.}.

\begin{figure}[hbt!]
\centering
\begin{subfigure}[t]{0.19\textwidth}
\centering
\includegraphics[width=.95\textwidth]{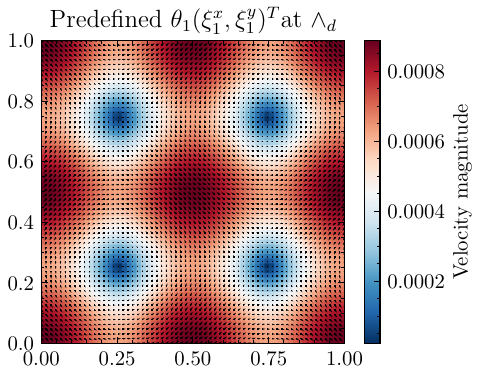}
\end{subfigure}
\begin{subfigure}[t]{0.19\textwidth}
\centering
\includegraphics[width=.95\textwidth]{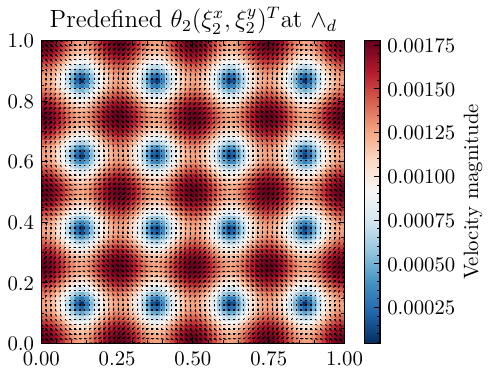}
\end{subfigure}
\begin{subfigure}[t]{0.19\textwidth}
\centering
\includegraphics[width=.95\textwidth]{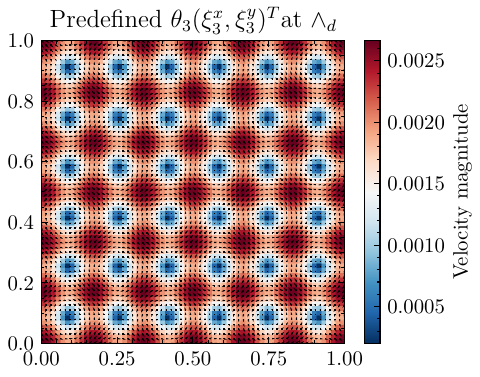}
\end{subfigure}
\begin{subfigure}[t]{0.19\textwidth}
\centering
\includegraphics[width=.95\textwidth]{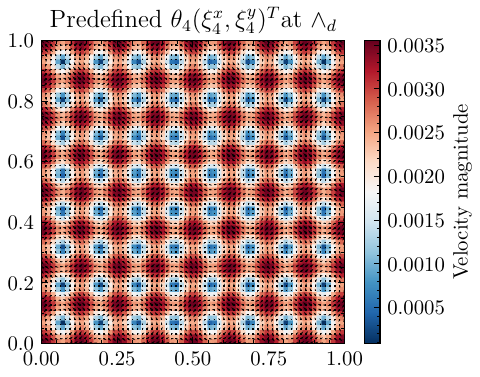}
\end{subfigure}
\begin{subfigure}[t]{0.19\textwidth}
\centering
\includegraphics[width=.95\textwidth]{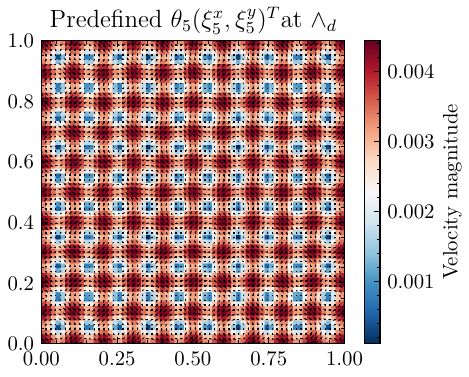}
\end{subfigure}
\\
\begin{subfigure}[t]{0.19\textwidth}
\centering
\includegraphics[width=.95\textwidth]{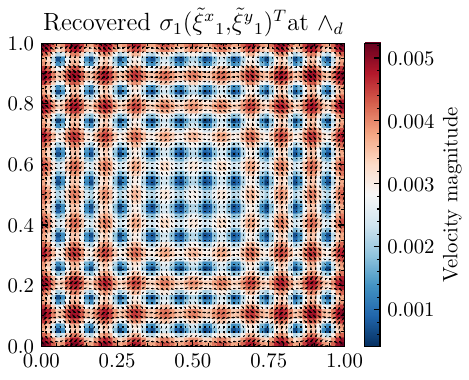}
\end{subfigure}
\begin{subfigure}[t]{0.19\textwidth}
\centering
\includegraphics[width=.95\textwidth]{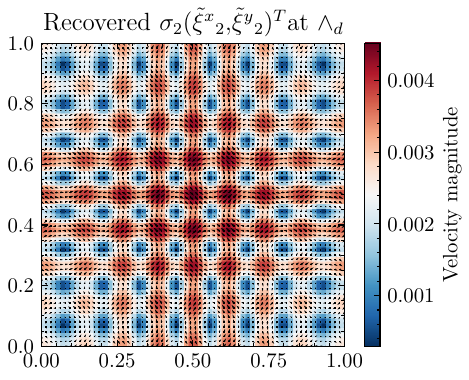}
\end{subfigure}
\begin{subfigure}[t]{0.19\textwidth}
\centering
\includegraphics[width=.95\textwidth]{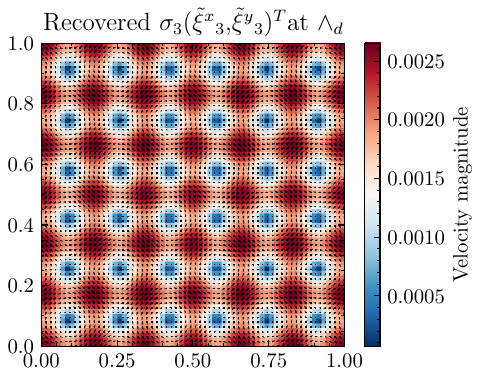}
\end{subfigure}
\begin{subfigure}[t]{0.19\textwidth}
\centering
\includegraphics[width=.95\textwidth]{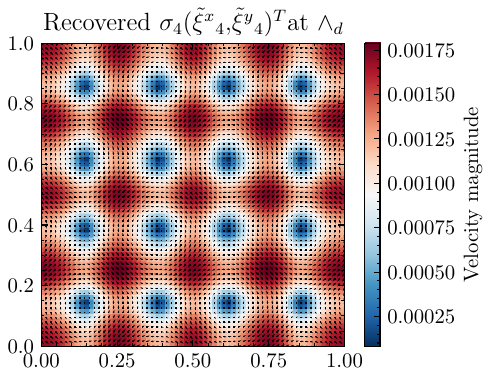}
\end{subfigure}
\begin{subfigure}[t]{0.19\textwidth}
\centering
\includegraphics[width=.95\textwidth]{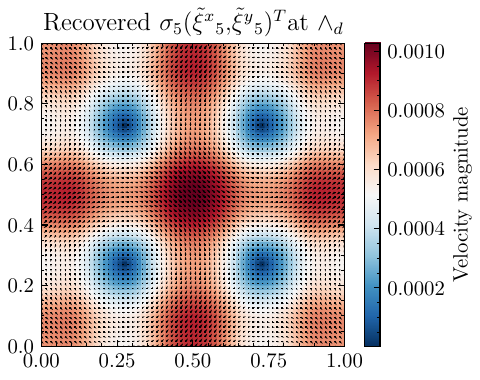}
\end{subfigure}
\caption{ 
The first row contains the vector fields and magnitude of the predefined basis $\lbrace\theta_{p}^{\ast}\xi_{p}\rbrace_{p=1}^{p=5}$ used in the generation of dataset 2, when evaluated at the weather stations $\b x_{\b i} \in \wedge_d$.  The second row shows $\lbrace \sigma_p (\tilde{\xi}^{x}_{p},\tilde{\xi}^{y}_{p})^T \rbrace_{p=1}^{p=5}$ the recovered basis from truncated SVD at weather stations, generated by the algorithm described in \cref{method:TSDVWSD}, not shown here is the 6th recovered principle component, a machine precision zero vector-field similar to \cref{fig:recovered sigma p2}.}
\label{fig: basis recovery p = 5}
\end{figure}

\begin{figure}[hbt!]
\centering
\begin{subfigure}[t]{0.325\textwidth}
\centering
\includegraphics[width=.95\textwidth]{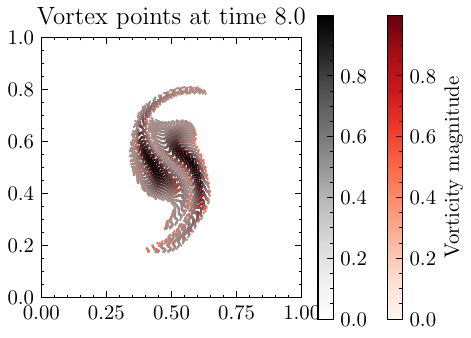}
\end{subfigure}
\begin{subfigure}[t]{0.325\textwidth}
\centering
\includegraphics[width=.95\textwidth]{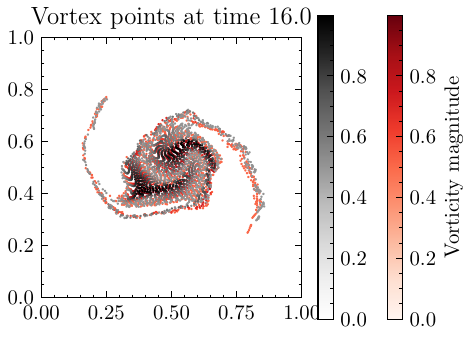}
\end{subfigure}
\begin{subfigure}[t]{0.325\textwidth}
\centering
\includegraphics[width=.95\textwidth]{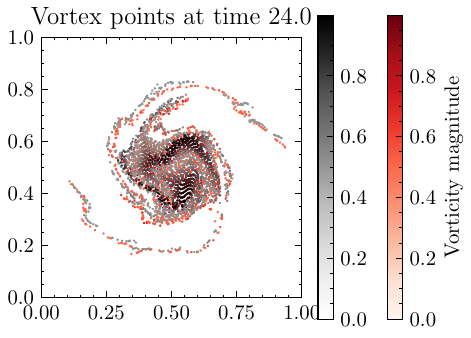}
\end{subfigure}\\
\begin{subfigure}[t]{0.325\textwidth}
\centering
\includegraphics[width=.95\textwidth]{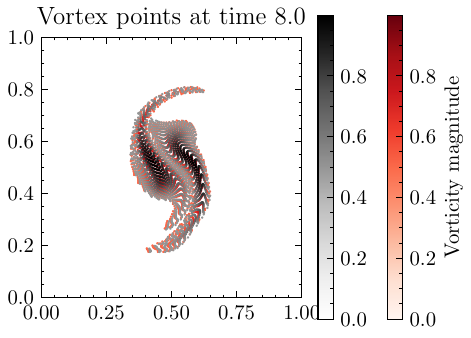}
\end{subfigure}
\begin{subfigure}[t]{0.325\textwidth}
\centering
\includegraphics[width=.95\textwidth]{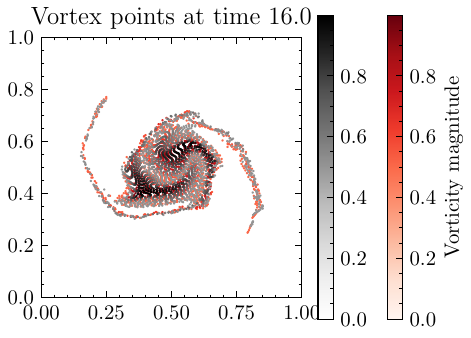}
\end{subfigure}
\begin{subfigure}[t]{0.325\textwidth}
\centering
\includegraphics[width=.95\textwidth]{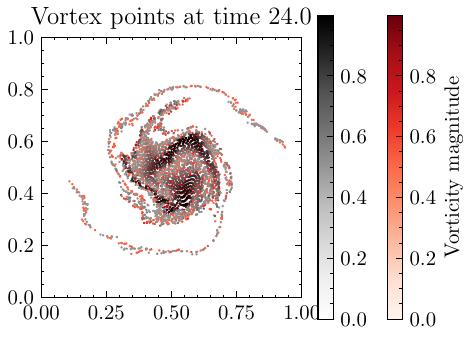}
\end{subfigure}\\
\caption{Dataset 2 $(P=5)$. In the first row, the red points are evolved using recovered increment and recovered basis functions from \cref{method:TSDVWSD} without the time mean contribution, and the grey points are the data. The second row contains the same experiment but with the additional time mean drift in.}\label{fig:dataset2 driven solution}
\end{figure}

In the first row of
\Cref{fig: basis recovery p = 5} we display the five basis functions used to generate the synthetic dataset 2, while in the second row, we plot the recovered 5 basis functions obtained through \cref{method:TSDVWSD}. Approximately 99.99 percent of covariance was explained by these five basis functions, with the sixth having machine precision magnitude. The recovered basis was not found unique or ordered (SVD is non-unique), but the vectorfields recovered exhibit some similarity in magnitude and shape to those used to generate dataset 2. \Cref{fig:dataset2 driven solution} contains the SPDE driven by the recovered signal $\lbrace \Delta W^r_{p} \rbrace_{p\in[5]}$ using the Fourier interpolated recovered 5 components $\lbrace\sigma_p \tilde{\b \xi_p}\rbrace_{p\in[5]}$, as compared with the original data. The relative $L^2$ error of the recovered increment time mean included driven solution from the data is $0.0531869$, the relative 
$L^2$ relative error of the recovered time mean not included solution is $0.0629821$. This indicates an improvement in the re-simulation of data by the inclusion of the time mean.

We test that recovered increments are normal. The Shapiro-Wilk test \cite{shapiro1965analysis} tests gives a score of 0.99863, and p-value of 0.421952. The two-sample goodness of fit Kolmogorov-Smirnov test (\cite{berger2014kolmogorov}) between $\Delta W^*$ and, $U/\Delta t= \Delta W^r$ gives a test statistic of 0.0390625, with p-value 0.990. This is not below the threshold of 0.05, so we cannot reject the null hypothesis that this sample is distributed according to the standard normal with confidence level 95 percent. We perform the Anderson test, the value of the test statistic was 0.317 and doesn't exceed the critical values [0.574, 0.654, 0.785, 0.915, 1.089], indicating the null hypothesis of normality cannot be rejected at the associated percent significance levels  $[15,  10,   5,   2.5 , 1. ]$. The recovered noise has moments displayed in the second column of \cref{table: moments}. Hypothesis testing indicates that the recovered increments are likely sampled from a normal distribution, justifying the basis $\lbrace \sigma_{p}\tilde{\b \xi}\rbrace_{p\in[P]}$ as a reasonable choice for modelling new normal increments.

\paragraph{Verification of learning} We observe that the CRPS loss on training data decreases drastically (initially), indicating improved forecast skill in comparison to a deterministic forecast for training dataset 2. The proposed initial parameter guess is $10^{-12}\b e_5$ (essentially a deterministic proposal), the synthetic data was generated with ``true'' parameter values $\b \theta^* = 10^{-5}\b e_5$. After 500 iterations of gradient descent, the estimated parameter values are 
 $[0.0001621, -0.00001877,-0.00002568, 0.00003455, 0.0001050]$ indicating identification of the rough sizes of the parameters, but not sign or exact size. We found examples of new parameters not equalling the true parameters used to generate the training data, but giving improved CRPS skill scores during training. 

\paragraph{Forecast Verification for underlying distribution}
For dataset 2, an additional 1000 hidden testing datasets were created, with the same $P=5$ basis functions with same parameter values $\b {\theta}^* = 10^{-5} \b{e}_{5}$, but with different driving increments. We tabulate the CRPS average scores in the second column of \cref{table:total}. We note that the backpropagation approach \cref{Method:ensemble_learning} had an unusually low CRPS average indicating good forecast skill. We tabulate the relative CRPSS scores in \cref{table:hidden datasets dataset 2 relative}. From which we conclude that in terms of relative CRPSS. The perfect model ensemble on average outperformed the Without-time mean ensemble by 1.0 percent. The Without-time-mean ensemble on average outperformed the With-time-mean ensemble by 2.5 percent. The With-time-mean ensemble outperformed the randomised initial condition ensemble by 2.8 percent. The randomised initial condition ensemble outperformed the Persistence ensemble by 81.5 percent. 


\begin{table}[htp]
\centering
\begin{tabular}{||c|c|c|c|c|c|c||}
\hline
scheme & RIC & With-time-mean & Without-time-mean & Perfect & Persistence \\
\hline 
\hline
RIC & 0 & -2.921 & -5.631 & -6.704 & 81.46 \\
\hline
With-time-mean & 2.838 & 0 & -2.634 & -3.676 & 81.98 \\
\hline
Without-time-mean & 5.331 & 2.566 & 0 & -1.016 & 82.45 \\
\hline
Perfect & 6.283 & 3.546 & 1.006 & 0 & 82.62 \\
\hline
Persistence & -439.3 & -455.1 & -469.7 & -475.5 & 0 \\
\hline
\end{tabular}
\caption{Hidden dataset 2, ``Percent improvement" table of row scheme over the column scheme for representing the distribution, i.e. hidden datasets. }
\label{table:hidden datasets dataset 2 relative}
\end{table}

From \cref{table:hidden datasets dataset 2 relative}, and \cref{fig:dataset2 driven solution} we conclude that in the instance when the data comes from a model without a time mean drift. Adding on the observed time mean drift was not helpful in representing the hidden model, despite predicting the training data more accurately. We speculate this occurs because the observed drift is a statistical error associated with sampling a unknown distribution rather than a systematic modelling error.

\subsubsection{Results: Dataset 3}

\begin{figure}[!hbt]
\centering
\begin{subfigure}[t]{0.31\textwidth}
\centering
\includegraphics[width=.95\textwidth]{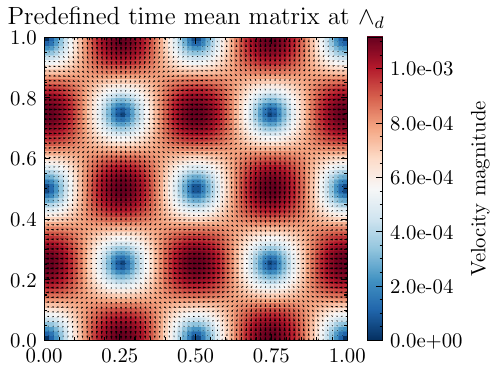}
\caption{\hfill}
\label{fig:predefined_itostrat}
\end{subfigure}
\begin{subfigure}[t]{0.3\textwidth}
\centering
\includegraphics[width=.95\textwidth]{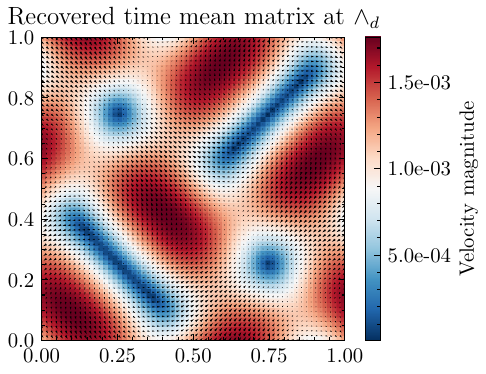}
\caption{\hfill}
\label{fig:recovered time mean itostrat}
\end{subfigure}
\begin{subfigure}[t]{0.3\textwidth}
\centering
\includegraphics[width=.95\textwidth]{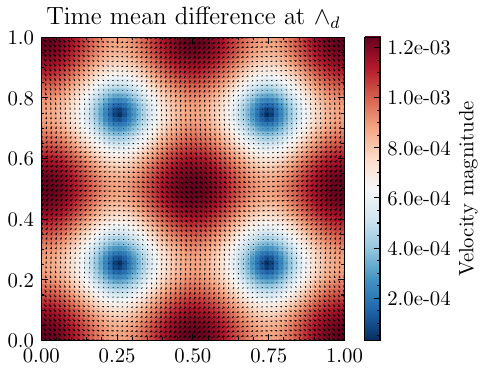}
\caption{\hfill}
\label{fig:time mean diffence itostrat}
\end{subfigure}
\caption{ Predefined Itô-Stratonovich drift \cref{fig:predefined_itostrat}(a), recovered time mean drift \cref{fig:recovered time mean itostrat}(b), attained by \cref{method:TSDVWSD} applied on dataset 3. The difference is plotted in \cref{fig:time mean diffence itostrat}(c), indicating the recovered time mean drift is a contribution from both the predefined difference from the proposed forward model and the data (Itô-Stratonovich correction) as well as the specific Brownian motion realisation associated with sampling from underlying hidden data distribution. }
\label{fig: recovery of mean in dataset 3}
\end{figure}

\paragraph{Twin experiment} Synthetic dataset 3 described in \cref{sec:synthetic data generation} was generated using a single basis of noise (\cref{eq: basis 1}) and a predefined drift \cref{eq: ito stratonovich drift function}(\cref{fig:predefined_itostrat}) mimicking unresolved small scale drift dynamics, the snapshot (at $t=0,16,32$) of vortex positions is shown in the third row of \cref{fig:total points evolving.}.

Using \cref{method:TSDVWSD}, we recover a time mean drift (plotted in \cref{fig:recovered time mean itostrat}), a basis function (not plotted indistinguishable to \cref{fig:recovered sigma p1}) and a driving signal, such that the DAM can be reconstructed to $1.90113e-13$. 
The recovered $\Delta W^r$ gives a Shapiro-Wilk test W score of 0.989228 with p-value 0.0538488, and a Shapiro statistic of 0.989228 with p value 0.0538488. The two-sample goodness of fit Kolmogorov-Smirnov test gives a KS statistic of 0.03515625, with p-value 0.997513 not below the threshold of 0.05. This indicates some evidence that the recovered increments are normal and the basis function is appropriate for use with different normal increments. 

\begin{figure}[hbt!]
\centering
\begin{subfigure}[t]{0.325\textwidth}
\centering
\includegraphics[width=.95\textwidth]{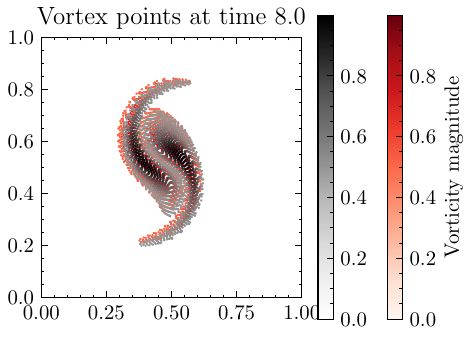}
\end{subfigure}
\begin{subfigure}[t]{0.325\textwidth}
\centering
\includegraphics[width=.95\textwidth]{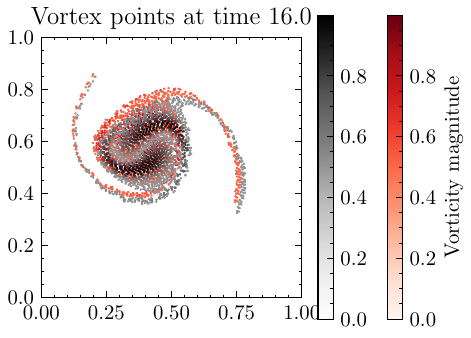}
\end{subfigure}
\begin{subfigure}[t]{0.325\textwidth}
\centering
\includegraphics[width=.95\textwidth]{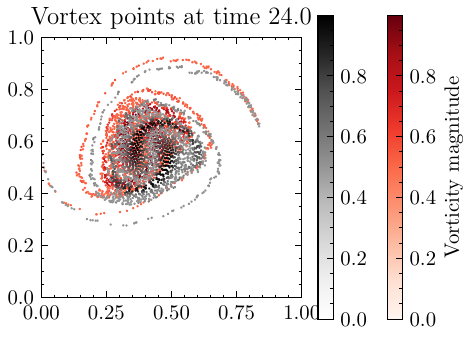}
\end{subfigure}\\
\begin{subfigure}[t]{0.325\textwidth}
\centering
\includegraphics[width=.95\textwidth]{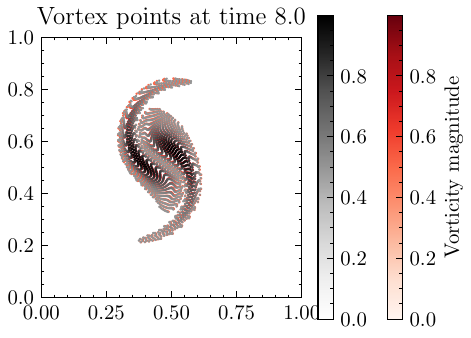}
\end{subfigure}
\begin{subfigure}[t]{0.325\textwidth}
\centering
\includegraphics[width=.95\textwidth]{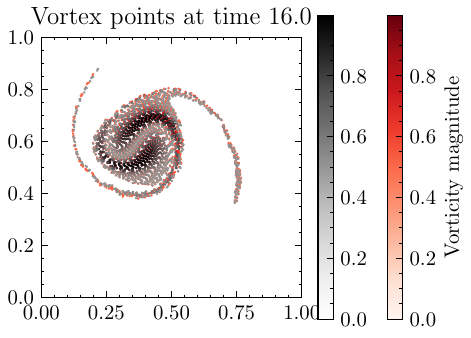}
\end{subfigure}
\begin{subfigure}[t]{0.325\textwidth}
\centering
\includegraphics[width=.95\textwidth]{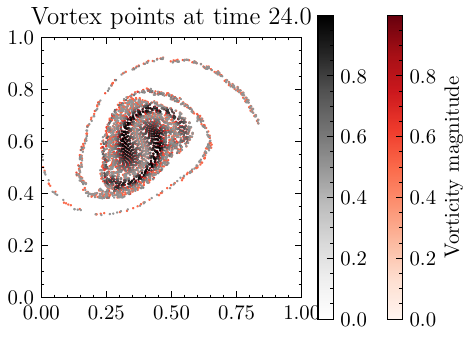}
\end{subfigure}
\caption{Dataset 3, reconstruction of data. Grey data points are the data. The red points are the recovered increment driven system. In the first row, the red points are evolved without using the additional time mean contribution using \cref{method:without}. The second row contains the same experiment but with the additional time mean drift using \cref{method:with}. We observe that the time mean velocity drift has a significant effect, in the context of this re-simulation of data in this twin experiment. } \label{fig:dataset 3 driven solution. }
\end{figure}

In \cref{fig:dataset 3 driven solution. } we plot the re-simulation of data with the recovered increments $\Delta W^r$, and the recovered basis function $\tilde{\b\xi}_1$ with and without the inclusion of the recovered time mean drift. The relative (spacetime L2) error of the time mean included solution is 0.0160130,
whereas the relative error of the recovered time mean not included solution is 0.0864620. We observe in both the relative L2 spacetime error and \cref{fig:dataset 3 driven solution. } that the recovered time mean is significantly helpful in the re-simulation of the observed dataset 3 training data. Demonstrating that the observed time mean drift can compensate for both the Itô-Stratonovich modelling deficiency and the unphysical drift observed in the DAM from statistical sampling error, in the re-simulation of data.

The interesting feature of dataset 3, is that the recovered drift (plotted in \cref{fig: recovery of mean in dataset 3}), is visibly affected by model inadequacy by missing an Ito-Stratonovich correction drift term \cref{fig:predefined_itostrat}, and also by statistical sampling error (associated with the specific Brownian motion realisation used for data). This is highlighted in \cref{fig:time mean diffence itostrat}, where the difference between the predefined time mean drift and recovered time mean drift, appears to be the same shape as the basis function \cref{eq: basis 1} for noise. This motivates computing the CRPSS on hidden data, to see whether the time mean drift is important in terms of representing the Itô-Stratonovich model deficiency despite the addition of an unphysical statistical drift bias observed by the data anomaly matrix. 

\paragraph{Forecast Verification for underlying distribution}
In the third column of \cref{table:total} we tabulate the averaged CRPS score over the hidden datasets. We turn this into the improvement in relative CRPSS displayed in \cref{table:Dataset3 relative improvement}. From this, we conclude that the Perfect ensemble model on average represented the hidden 1000 datasets 0.9 percent better than the Without-time-mean ensemble. The Without-time-mean ensemble on average represented the 1000 hidden datasets  1.1 percent better than the With-time-mean ensemble. The With-time-mean ensemble on average represented the 1000 hidden datasets 9.3 percent better than the RIC ensemble. The RIC ensemble outperformed the persistence ensemble by 47 percent. 

We conclude that the inclusion of a drift term from data, even when well motivated from modelling deficiencies may not necessarily improve the calibrated model in terms of CRPS score. We speculate (based on the previous two experiments) that this occurs because the finite realisation of the $\xi dW$ term in the generation of the synthetic training data resulted in an unphysical time mean drift observed in the DAM. Whose inclusion in a calibrated model with drift (e.g. \cref{method:with}) can dominate the potential benefit of modelling a small but well-motivated drift.


\begin{table}[!htp]
    \centering
\begin{tabular}{||c|c|c|c|c|c|c||}
\hline
scheme & RIC & With-time-mean & Without-time-mean & Perfect & Persistence \\
\hline\hline
RIC & 0 & -10.29 & -11.51 & -12.52 & 47.9 \\
\hline
With-time-mean & 9.326 & 0 & -1.11 & -2.029 & 52.76 \\
\hline
Without-time-mean & 10.32 & 1.098 & 0 & -0.9087 & 53.28 \\
\hline
Perfect & 11.13 & 1.989 & 0.9005 & 0 & 53.7 \\
\hline
Persistence & -91.96 & -111.7 & -114.1 & -116 & 0 \\
\hline
\end{tabular}
   \caption{Hidden dataset 3, ``Percent improvement" table of row scheme over the column scheme for representing the distribution, i.e. hidden datasets.}
\label{table:Dataset3 relative improvement}
\end{table}

\subsubsection{Results: Dataset 4}

\begin{figure}[hbt!]
\centering
\begin{subfigure}[t]{0.31\textwidth}
\centering
\includegraphics[width=.95\textwidth]{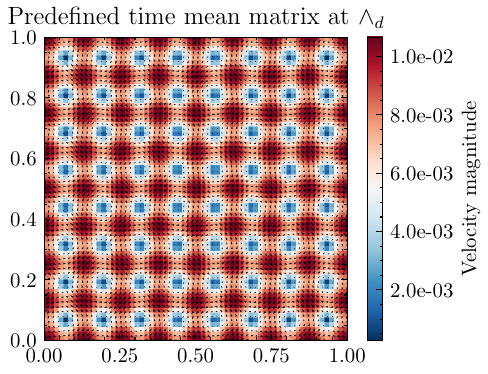}
\caption{\hfill}
\label{fig:predefined_timemean_physical}
\end{subfigure}
\begin{subfigure}[t]{0.3\textwidth}
\centering
\includegraphics[width=.95\textwidth]{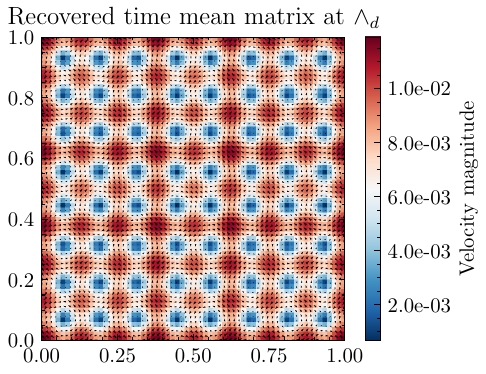}
\caption{\hfill}
\label{fig:recovered time mean matrix_physical}
\end{subfigure}
\begin{subfigure}[t]{0.3\textwidth}
\centering
\includegraphics[width=.95\textwidth]{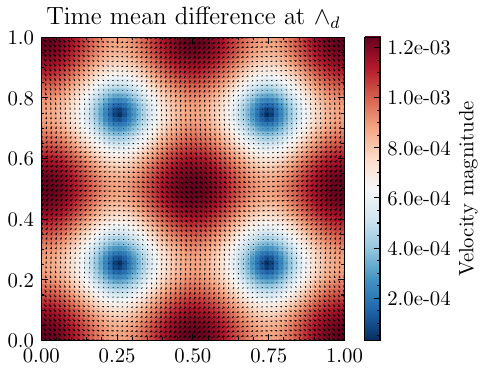}
\caption{\hfill}
\label{fig:time mean diffence_physical}
\end{subfigure}
\caption{ Predefined time mean drift \cref{fig:predefined_timemean_physical}(a), recovered time mean drift \cref{fig:recovered time mean matrix_physical}(b), after
method \cref{method:TSDVWSD} is applied on dataset 4. The difference is plotted in \cref{fig:time mean diffence_physical}(c), indicating the recovered time mean matrix is a contribution from both the physical predefined difference from the proposed forward model and specific Brownian motion realisation associated with sampling from the underlying hidden data distribution. }
\label{fig: recovery of mean in dataset 4}
\end{figure}

\begin{figure}[hbt!]
\centering
\begin{subfigure}[t]{0.325\textwidth}
\centering
\includegraphics[width=.95\textwidth]{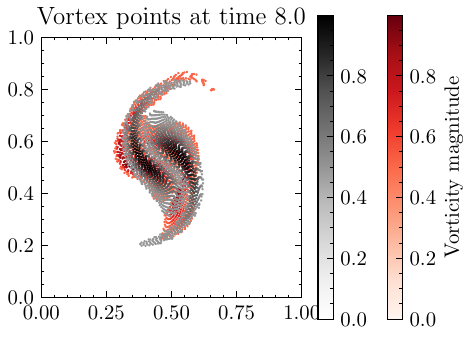}
\end{subfigure}
\begin{subfigure}[t]{0.325\textwidth}
\centering
\includegraphics[width=.95\textwidth]{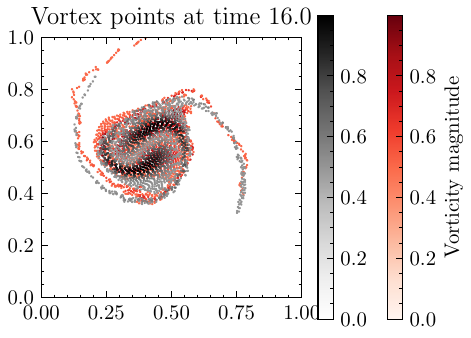}
\end{subfigure}
\begin{subfigure}[t]{0.325\textwidth}
\centering
\includegraphics[width=.95\textwidth]{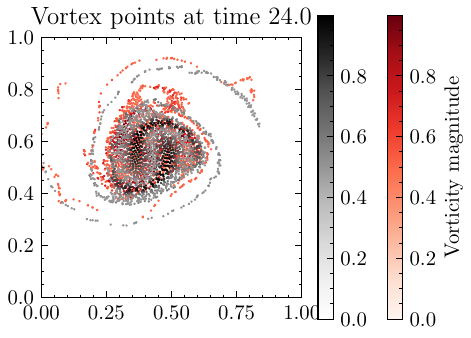}
\end{subfigure}\\
\begin{subfigure}[t]{0.325\textwidth}
\centering
\includegraphics[width=.95\textwidth]{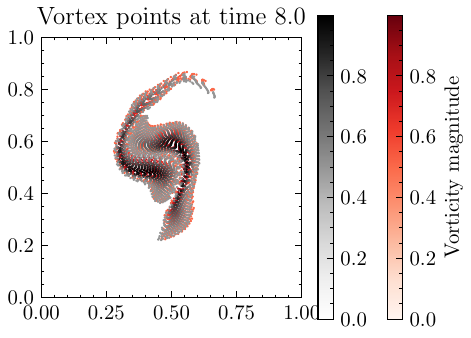}
\end{subfigure}
\begin{subfigure}[t]{0.325\textwidth}
\centering
\includegraphics[width=.95\textwidth]{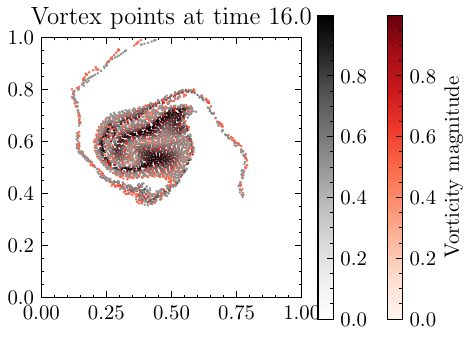}
\end{subfigure}
\begin{subfigure}[t]{0.325\textwidth}
\centering
\includegraphics[width=.95\textwidth]{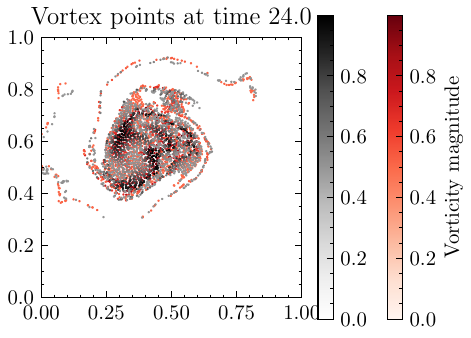}
\end{subfigure}
\caption{Dataset 4, reconstruction of data. Grey data points are the data. The red points are the recovered increment driven system. In the first row, the red points are evolved without using the additional time mean contribution using \cref{method:without}. The second row contains the same experiment but with the additional time mean drift using \cref{method:with}. We observe that the time mean velocity drift has significant effect, in the context of this re-simulation of data in this twin experiment. } \label{fig:dataset four driven solution. }
\end{figure}

\paragraph{Twin experiment}
Synthetic dataset 4 described in \cref{sec:synthetic data generation} was generated using a single basis of noise (\cref{eq: basis 1}) and a predefined drift \cref{eq: physical drift}(\cref{fig:predefined_timemean_physical}) mimicking the effect of unresolved small scale drift dynamics, the snapshot (at $t=0,16,32$) of vortex positions is shown in the fourth row of \cref{fig:total points evolving.}. 

Using \cref{method:TSDVWSD}, we recover a time mean drift (plotted in \cref{fig:recovered time mean matrix_physical}), a basis function and a driving signal, such that the DAM can be reconstructed to 1.98225e-13. On the recovered increments we perform the Shapiro-Wilk test to evaluate the null hypothesis that the data was drawn from a normal distribution and get a score of 0.989228, and p-value of 0.0538488.
We perform the two-sample goodness of fit Kolmogorov-Smirnov test for the recovered increments to test the null hypothesis that the recovered increments are distributed according to the appropriately scaled normal distribution. The p-value of 0.997513 is not below the threshold of 0.05, so we cannot reject the null hypothesis that this sample is distributed according to the standard normal with a confidence level 95 percent. Giving evidence that $\tilde{\b\xi}_1$ is an appropriate basis for stochastic parametrisation.

The interesting feature of dataset 4 is that the recovered time mean drift is made up from both real model inadequacies from missing a physical drift term in the underlying model (plotted in \cref{fig:predefined_timemean_physical}) and specific sampling error. This is highlighted in \cref{fig:time mean diffence_physical}, where the difference between the recovered and predefined drift is plotted and appears to be in the same shape as the stochastic basis velocity \cref{eq: basis 1}.

Snapshots (at t=8,16,24) of the re-simulation of data with and without the time mean drift are plotted in the first and second row of \cref{fig:dataset four driven solution. } respectively. The relative spacetime error of re-simulating data with the time mean included is 0.0672719 whereas
the relative error without the time mean included solution is 0.214116. We conclude that the inclusion of the time mean drift term is helpful in the re-simulation of the training dataset 4. Since this model-data mismatch in drift is larger in magnitude than the Itô-Stratonovich correction, it is well motivated to consider whether the observed time mean drift can be used in an ensemble forecast and improve the forecast skill. Despite the potential for a statistical bias associated with the sampling of the Brownian motion.

\paragraph{Forecast Verification of underlying distribution}
In the fourth column of \cref{table:total} we tabulate the averaged CRPS score of each model over the hidden data. This is presented in terms of a relative improvement in average CRPSS in \cref{table:Dataset four relative averaged skill score table for representing the hidden data sets.}. We conclude that in terms of representing the 1000 hidden realisations from the underlying SPDE/SDE, as compared by relative space-time averaged CRPSS. The RIC ensemble was on average 42.41 percent better than Persistence. Without-time-mean was on average 11.08 percent better than RIC. With-time-mean was on average 3.849 percent better than Without-time-mean. The Perfect model was on average 1.745 percent better than With-time-mean. 

We make the following important conclusion from dataset 4. If the data is generated with a model with a notable time mean drift, using \cref{method:TSDVWSD} the time mean can be captured. Furthermore the inclusion of the measured drift (\cref{method:with}) improved the re-simulation of data as in \cref{fig:dataset four driven solution. }. The inclusion of the measured drift also improved the average CRPS of hidden testing datasets as seen in \cref{table:total,table:Dataset four relative averaged skill score table for representing the hidden data sets.}, even in the presence of statistical sampling error. 

We hypothesise that observed time mean drifts are likely to be significant and expected in realistic modelling scenarios, there are likely unresolved drift processes between the forward model and observed data. In which case including the observed time mean as in \cref{method:with}, can be seen as an essential modelling step unless one has access to diagnostics tools capable of ruling out the data observed drift as a modelling error and classifying it as a statistical error. Such situations are unlikely, one does not necessarily get to resample from the underlying distribution from which the data was generated as we have in this idealised testing scenario.


\begin{table}[htp]
\centering
\begin{tabular}{||c|c|c|c|c|c|c||}
\hline
scheme & RIC & With-time-mean & Without-time-mean & Perfect & Persistence \\
\hline\hline
RIC & 0 & -16.96 & -12.46 & -19.04 & $\b{42.41}$ \\ 
\hline
With-time-mean & 14.5 & 0 & $\b{3.849}$ & -1.776 & 50.76 \\ 
\hline
Without-time-mean & $\b{11.08}$ & -4.003 & 0 & -5.85 & 48.79 \\ 
\hline
Perfect & 15.99 & $\b{1.745}$ & 5.527 & 0 & 51.62 \\ \hline
Persistence & -73.64 & -103.1 & -95.28 & -106.7 & 0 \\
\hline
\end{tabular}
\caption{Hidden dataset 4, ``Percent improvement" table of row scheme over the column scheme for representing the distribution, i.e. hidden datasets.}
\label{table:Dataset four relative averaged skill score table for representing the hidden data sets.}
\end{table}


\subsection{Summary of results}

\begin{table}[!htp]
\centering
\begin{tabular}{||c||c||c||c||c||}
\hline
scheme & Average CRPS & Average CRPS & Average CRPS & Average CRPS \\
& Dataset 1.& Dataset 2.& Dataset 3.& Dataset 4.\\
\hline\hline
Persistence & 1.501e-01& 1.223e-01& 1.509e-01 & 1.520e-01 \\
\hline
RIC & 7.591e-02& 2.268e-02& 7.859e-02 & 8.755e-02 \\
\hline
Perfect & 6.836e-02& 2.126e-02& 6.985e-02& 7.354e-02 \\
\hline
Without-time-mean  & 6.834e-02& 2.147e-02& 7.048e-02 & 7.785e-02 \\
\hline
With-time-mean & 6.979e-02 &2.204e-02& 7.126e-02&  7.485e-02 \\
\hline
 Learned forward model: $E_{o}=5$ & 6.944e-02& 1.789e-02 &NA&NA\\
\hline
\end{tabular}
\caption{Each column contains the average CRPS scores associated with attaining hidden test datasets, from an ensemble(denoted in a row) calibrated on the corresponding training dataset. }
\label{table:total}
\end{table}


Under the assumption that the data comes from an SPDE realisation, the new SVD calibration technique \cref{method:TSDVWSD} captures the same number of basis functions used to generate the data for $P=1$ and for $P=5$. The basis recovered are in some objectionable sense reasonable in magnitude and shape in comparison to the basis used to generate the data. The recovered noise increments have passed several hypothesis tests indicating normality. In the specific instance of one basis function, the recovered basis function was shown to be in agreement with the basis function used to generate the data, both visually and agreeing to $10^{-2}$ in the relative error norm. 

We speculated that the $10^{-2}$ relative $L^2$ discrepancy in both path and basis function in part came from the time mean removal in the SVD decomposition, and proposed adding this term back in as a deterministic drift velocity in the equation without violating the geometric structure of the model. To test the addition of this term we proposed the resimulation of data, by driving the SPDE with recovered increments, and recovered basis functions. In the resimulation of data, driving the solution of the model could more accurately represent the training dataset by the inclusion of the time mean drift velocity, for all datasets. 

We also estimated the Continuous Rank Probability Score (CRPS) for each model, the persistence forecast (\cref{method: persistence}), the new SVD algorithm with the time mean (\cref{method:with}), and without it (\cref{method:without}), the perfect model (\cref{method: perfect ensemble}), the and the model rerun with learned parameter values (\cref{Method:ensemble_learning}). This was done by estimating CRPS over all time and all state space values for a 30-member ensemble forming a global estimate of the CRPS score to quantify how likely the observations come from the ensemble. This is performed on 1000 hidden datasets and averaged to help distinguish the sampling error associated with sampling from the data distribution.

From which we concluded. Should data come from an SPDE/SDE realisation with a physical drift term, using the time mean from the DAM to evolve the ensemble is an important modelling step for improving the skill of the forecast. The potential drawback of adding a time mean drift term is that, if the data arises from a SPDE/SDE realisation with a small or insignificant drift term, sampling from the data distribution may result in an observed non-physical drift in the DAM matrix. The appearance of a nonphysical drift is typically small in magnitude (arising from the statistical error of sampling Brownian motion not having mean zero) and may justify ignoring small drifts such as the higher order (typically smaller) Itô-Stratonovich correction in the context of calibration. However, the possibility of a nonphysical drift in the DAM does not justify neglecting a drift of larger magnitude arising from model-data mismatches. Overall both SVD approaches \cref{method:with} and \cref{method:without} (with and without the time mean) outperformed both RIC and Persistence, and approached the CRPS score of the perfect model.

Regarding \cref{method:backprop ensemble learning stage}, evidence points towards a decreasing CRPS score with increased training time, and showed a 40 percent CRPSS improvement from the initial deterministic proposed ensemble. The lack of interpolation lead to a drastic improvement in compute speed in the forward model. With a fixed number of basis functions the SVD approaches produced ensemble forward models that took approximately 1 hour to run, whilst the parameter estimated model took approximately 1 minute to run, this scaling gets more drastic with the more weather-stations used in the model. It is also worth remarking the backpropagation approach \cref{method:backprop ensemble learning stage} did not use any Eulerian weather station data in the training, and only trained on approximately 3 percent of available data Lagrangian path data, so direct comparison to SVD methodology may not be appropriate.

There are many free parameters involved, $n_c,E_o,T_o,n_v,P,n_{t}...$, in which the effectiveness of both calibration methods could be studied. For example, we plot the CRPSS and parameter magnitude estimation error for $E_o = \lbrace 1,10,100 \rbrace$, trained over the entire time window $\mathcal{T}_{o} = [256]/32$ in
\cref{fig:training on more ensemble members.}, for a smaller dataset to illustrate the effect of $E_o$ size during training. Back-propagation through an ensemble costs vastly more than an SVD approach. Nevertheless, in the context of offline approaches to data assimilation, it may be beneficial to incur the cost of the offline training, to improve the forward ensemble model speed (due to the lack of interpolation in the forward model).

\section{Conclusion and future outlook} \label{sec: conclusion}

A methodology for the calibration of stochastic vector-fields was proposed using the combination of the Biot-Savart kernel and vortex positional data in combination with a SVD approach \cref{method:TSDVWSD}. The Eulerian vector-fields recovered are demonstrated relevant to the stochastic forward model proposed and we provided evidence \cref{method:TSDVWSD} parameterised the difference between the proposed forward model and the synthetic data. The methodology has been shown consistent by the use of a twin experiment and using the CRPS skill score as compared relative to several benchmarks including the ``perfect" model (\cref{method: perfect ensemble}). 

The inclusion of a time mean drift velocity was motivated and shown important in the context of a twin experiment. This term was shown consistent with the geometric modelling assumptions (Kelvin theorem, coadjoint action \cref{sec:geometric interpretation}, Hamiltonian and Poisson structure \cref{sec:hamiltonian and poisson structure}). The forecast verification results presented here suggest the addition of the time mean drift term is an advantageous modelling step, except in the setting when the data does not have a large difference in drift from the model. Realistic data may have a significantly larger time mean drift velocity from the proposed forward model than in the idealised experiments presented here, making this arguably an essential modelling step.

\begin{figure}[hbt!]
\centering
\begin{subfigure}[t]{0.32\textwidth}
\centering
\includegraphics[width=.95\textwidth]{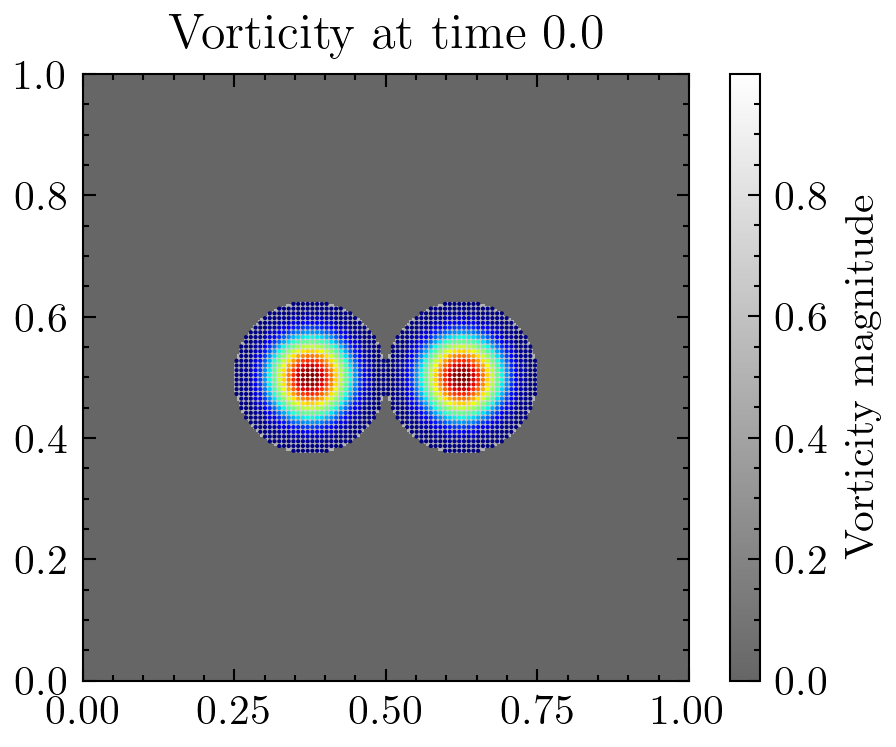}
\end{subfigure}
\begin{subfigure}[t]{0.32\textwidth}
\centering
\includegraphics[width=.95\textwidth]{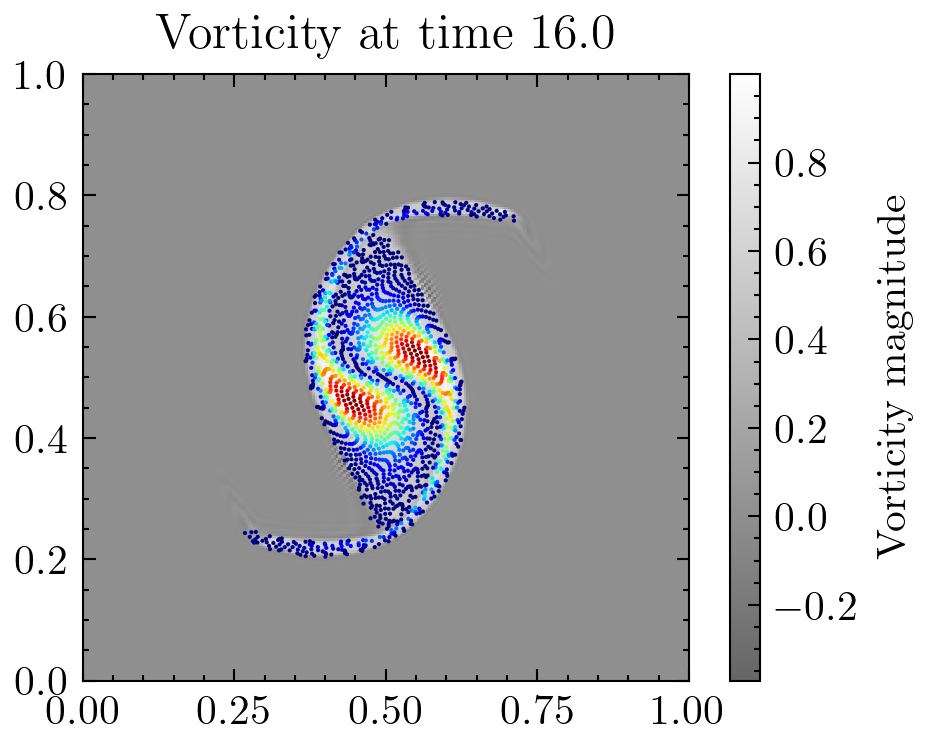}
\end{subfigure}
\begin{subfigure}[t]{0.32\textwidth}
\centering
\includegraphics[width=.95\textwidth]{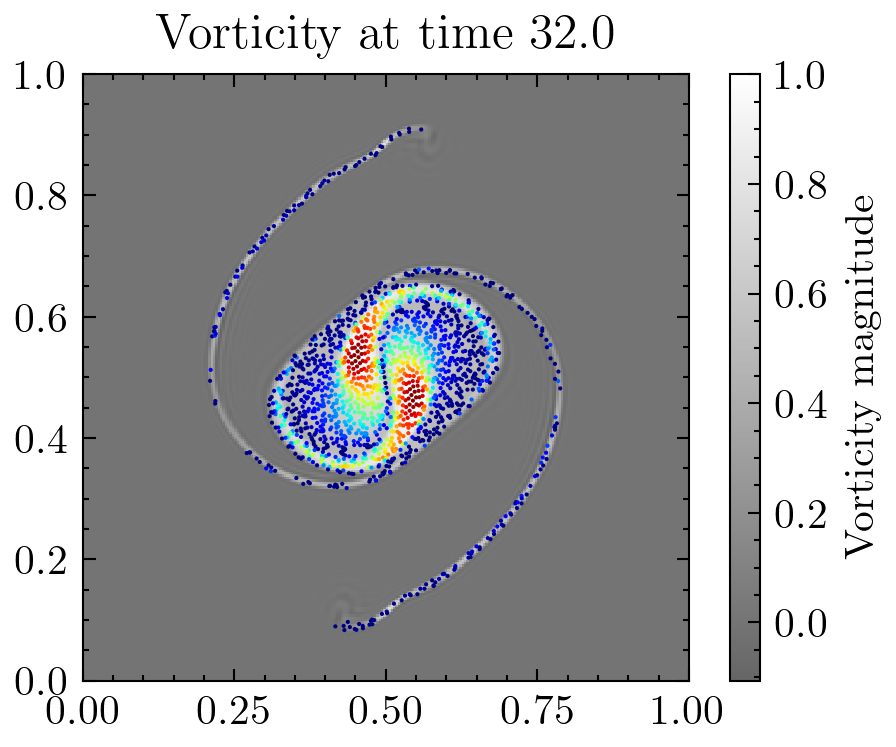}
\end{subfigure}
\caption{ Passive tracers carried by the Fourier reconstructed velocity field within a unlimited upwind bias high order flux form mimetic finite volume method (plotted grey) (solving strategy uses ideas from \cite{arakawa1981potential,woodfield2023new,hundsdorfer1995positive,gholami2016fft}), the passive particle tracer are coloured by the initial vorticity (plotted jet).}
\label{fig:finitevol}
\end{figure}

Using the Biot-Savart kernel to create a data anomaly matrix is not necessarily restricted to the setting of inviscid vortex methods, using the Biot-Savart kernel is likely applicable for the calibration of other fluid mechanics models. For the point of testing and to distinguish between model error data sample error and calibration error, we have biased all tests towards SPDE/SDE realisation data for which we know the parameters. The application of the calibration methodology proposed in this paper could be adapted in the context of less idealised (and less testable) synthetic data. For example, the vorticity equation can be solved (\cref{eq:2d salt euler}, \cref{example: 2d Euler}), using a finite volume discretisation, with passive tracer drifters (with an initially measured known vorticity) in the flow shown in \cref{fig:finitevol}. These tracers could be treated as analogues of the point vortices in the inviscid vortex method, and vector fields could be calibrated using \cref{method:TSDVWSD}. One could equally treat this as a reference dataset to calibrate an inviscid vortex method, serving as another example of stochastic coarse-grained model reduction.

Using a tangent linear model approach to auto differentiate through the ensemble forecast, as to minimise the CRPS Skill score \cref{method:backprop ensemble learning stage}. We were able to consistently propose basis functions with an improved relative CRPS skill score as compared to a near deterministic forecast, and did not require weatherstation data. The minimisation of the CRPS of an ensemble forecast did go someway to propose reasonable estimation of the true parameters magnitude (better than the initial guess). However, we found cases in which a decreases in CRPS did not necessarily lead to a more accurate estimation of parameter values. Parameters converged to roughly the correct magnitude of the proposed parameters, but not the right sign or exact same number.

The inconsistency in the calibration problem as posed as minimisation of a CRPS score and as a parameter estimation problem, could be a barrier for reliable methodology but also a potential opportunity for model-specific calibration. One could foresee calibration vector fields chosen to produce a good (in a probabilistic sense) forecast of a specific event. Providing an event-informed approach to the choice of vector fields. One could potentially achieve such a task using an approach similar to \cref{method:backprop ensemble learning stage} but with the minimisation of a Brier skill score of a single observed important event, e.g. hurricane hitting a specific location, sea surface height under a satellite track.

\paragraph{Acknowledgements.}
I would like to acknowledge Darryl Holm for continued support and insight. I would like to acknowledge Wei Pan, Ruiao Hu for valuable insights into existing calibration methodology. I would like to acknowledge Theo Diamantakis and Darryl Holm regarding various geometric insights into point vortices and the SALT framework in the variational principle. I would like to acknowledge Wei Pan, Oliver Street, Alex Lobbe for interesting discussions in weather forecast verification techniques. I would like to acknowledge discussions regarding numerical methods with Ruiao Hu, Wei Pan as well as Aythami Bethencourt de Leon, and James Micheal Leahy regarding specifics in the JAX coding environment. I would also like to acknowledge two particularly helpful and thorough anonymous reviewers for comments leading to the improvement of this manuscript.

The work of JW is supported by the European Research Council (ERC) Synergy grant “Stochastic Transport in Upper Ocean Dynamics” (STUOD) – DLV-856408.

\bibliographystyle{plainurl}
\bibliography{refs.bib}
\begin{appendices}

\section{Geometric structure}\label{sec:geometric interpretation}
It may be worth noting that although \cref{eq: time mean equation} appears to have an additional drift not present in the original Stochastic Advection by Lie Transport(SALT) paper \cite{holm2015variational}. The modelling remains faithful to the geometric framework in the following way. An additional deterministic drift velocity acting at the level of the particle trajectory map, gives a stochastic Kelvin theorem akin to \cite{holm2015variational} but with the additional feature of a data informed deterministic drift moving the loop \begin{align}
d\oint_{C(\b u dt +  \bar{\b v}dt + \xi_p \circ  dW^P)}\b u \cdot d \b x =0.
\end{align} An Euler Poincaré equation(see \cite{holm1998euler}) with a modification to the Lie algebra resulting in a coadjoint operator of the form 
$\operatorname{ad}^*_{\b u dt +  \bar{\b v}dt + \xi_p \circ dW^P}.$
In the Euler equation an additional deterministic drift term appears in the velocity of the Lie derivative operator, in 2D this appears as additional transport of vorticity by a time mean drift velocity as follows
\begin{align}
    d \omega_t + (\b u + \bar{\b v} \cdot \nabla )\omega_t dt + \sum_{p=1}^{P} (\theta_p \b \xi_{p}\cdot \nabla )\omega_t \circ dW^{p}=0.
\end{align}

\section{Hamiltonian and Poisson structure}\label{sec:hamiltonian and poisson structure}

The finite dimensional regularised Stratonovich system has a ``Hamiltonian" structure, when the vectorfield basis is assumed to have a streamfunction representation $\xi_p = -\nabla^{\perp}\psi_p, \quad \forall p\in [P]$, and when the time mean drift has a streamfunction representation $\bar{\b v} = -\nabla^{\perp}\bar{\psi} $. We first define the operators $\nabla^{\perp}_{\b i} := (-\partial_{y_{\b i}},\partial_{x_{\b i}})$, $\nabla_{\b i} := (\partial_{x_{\b i}},\partial_{y_{\b i}})$, denoting derivatives with respect to specific particle positions. Then the Hamiltonian system can be written as the following
\begin{align}
\Gamma_{\b i} d \b x_{\b i} = -\nabla^{\perp}_{\b i} \left(H_{\delta} dt + \sum_{\b i \in \wedge_0} \Gamma_{\b i} \bar{\psi}(\b x_{\b i})dt + \sum_{\b i \in \wedge_0}\sum_{p=1}^{P}\Gamma_{\b i} \theta_p \psi_{p}(\b x_{\b i})\circ dW^{p} \right), \quad \forall \b i \in \wedge_{0},
\label{eq:PV_SDE}\end{align}
 where $H_{\delta}$ is the deterministic regularised Kirchhoff Hamiltonian  
 \begin{align}
     H_{\delta} = \sum_{\b i, \b j\in \wedge^{\b i}_0, \b i\neq \b j}\Gamma_{\b i}\Gamma_{\b j}G_{\delta}(\b x_{\b i} - \b x_{\b j}).\label{eq:KirchoffHamiltonian}
 \end{align}

For a function of particle positions $F(t, \lbrace \b x_{\b i} \rbrace_{\b i \in \wedge_0^{\b i}})$, the time derivative reveals the following Poisson structure
\begin{align}
0 
&=
\partial_t F + \lbrace F,H_{\delta} \rbrace dt + \lbrace F,\bar{\psi} \rbrace dt +\sum_{p=1}^{P} \lbrace F, \theta_p\psi_{p} \rbrace 
\circ dW^p, \quad 
\lbrace F,H_{\delta} \rbrace := \sum_{\b i \in \wedge^{\b i}_0} \frac{1}{\Gamma_{\b i}}(\nabla_{\b i} F )\cdot (-\nabla_{\b i}^{\perp} H_{\delta}). \label{eq:poisson bracket}
\end{align}

\begin{example}[Surface Quasi-Geostrophic]\label{example: 2d SQG}
Surface Quasi-Geostrophic on $\mathbb{R}^2$ has the following differential relationship between deteministic stream function and vorticity, $\psi = 	(-\Delta)^{-1/2}\omega$, with Greens function given by $G(\b x)=-(2\pi)^{-1}||\b x||_2^{-1}$, and kernel by $K(\b x) =(4\pi)^{-1} \bx^{\perp} ||\b x||^{-3}$. This two dimensional model bears analogy to the three dimensional Euler Equation and kernel.  
\end{example}

\begin{example}[Quasi-Geostrophic Shallow Water]\label{example: 2d QGSW}
Quasi-Geostrophic Shallow Water on $\mathbb{R}^2$ has the following differential relationship between deterministic stream function and vorticity, $\psi = 	(\Delta - \lambda^2)^{-1}\omega$, with Greens function given by $G(\b x)=-(2\pi)^{-1}K_{0}(\lambda||\bx ||)$, and kernel by $K(\b x) =\lambda(2\pi)^{-1}K_{1}(\lambda||\bx||)$, 
where $K_n$ is the modified Bessel function of second kind, and satisfies $K_0' = -K_1$. Numerically, the modified Bessel function of the second kind requires approximation, typically done via numerical integration either through trapesium rule \cite{holm2006euler} or peicewise Chebyshev quadrature, additional computational time or computational resources are required.
\end{example}

\begin{example}[Euler-$\alpha$ (model of turbulence)]\label{example: 2d Euler alpha}
Euler-$\alpha$ (model of turbulence)on $\mathbb{R}^2$ has the following differential relationship $
\psi = 	(1-\alpha^2\Delta)^{-1} (-\Delta  )^{-1}\omega$. Using the fundamental solutions to the Helmholtz and Laplace operator D. D. Holm, M. Nitsche and V. Putkaradze \cite{holm2006euler} deduce the following Greens function $G(\b x) = \frac{-1}{2\pi} (\log (||\b x||_2) + K_{0}(||\b x||_2/\alpha))$ and Biot-Savart kernel
$ K(\b x) =(2\pi)^{-1} \bx^{\perp} ||\b x||_2^{-2}(1- ||\b x||_2/\alpha K_1(||\b x||_2/\alpha) )$. Here one observes a regularisation of the Euler kernel, taking a similar but distinct form to that of the regularised vortex blob approximations. There is exponential decay for large values of $||\b x||_2$, but unlike the vortex blob method the Greens function remains unbounded at the origin. The reconstructed velocity is bounded, but the vorticity is not. 
\end{example}

\begin{remark}[Different Domains.]
    Various modifications to the Greens functions have been proposed to deal with different domains. Point vortices have been proposed on a singularly periodic strip \cite{rosenhead1931formation, krasny1986study} using the method of images, a doubly periodic square see \cite{weiss1991nonergodicity} \cite{o1989hamiltonian}, for toroidal see \cite{sakajo2016point}, for gaps in walls see \cite{crowdy2006motion}, for inclusion of islands see \cite{johnson2005point}.  Milne-Thompson theorem for a cylinder can be used to take into account the introduction of a cylinder \cite{chorin1973numerical}, and the conformal Schwarz–Christoffel mapping could be proposed more generally. Embedded closed surfaces conformal to the unit sphere are considered in \cite{dritschel2015motion}, and for additional literature review see \cite{ball2006point} \cite{aref1988point}.
\end{remark}

\begin{remark}[CRPS]\label{remark:crps:notes}
The CRPS 
is a continuous version of the Rank probability score, and can be interpreted as a Brier score but integrated over all possible thresholds. The CRPS is a strictly proper scoring rule \cite{matheson1976scoring}. The CRPS measures differences between the predicted and occurred cumulative density functions and gives a score indicating how good the ensemble is at matching observations. 0 is accurate, and 1 is inaccurate. For a deterministic forecast, the CRPS reduces to the mean absolute error.
We use an estimator of the representation (pg8 eq 11 in \cite{gneiting2011comparing}) in the loss function with an additional fairness modification see \cite{ferro2014fair,zamo2018estimation}, see same reference for the equivalence to other formulations of the CRPS. For discussion, literature review and details regarding the Continuous Rank Probability Score, and relationship to other skill scores (such as Breir skill score \cite{brier1950verification}, and RPS) and other diagnostic tools such as (Rank Histogram/Talagrand diagram) see \cite{hersbach2000decomposition}, \cite{gneiting2011comparing}, and \cite{gneiting2007strictly}. For further insights into the Candille–Talagrand (2005), Brier score, Quantile score,  Hersbach and more decompositions of the CRPS score see \cite{arnold2023decompositions}. In the context of postprocessing of ensemble forecasting in the forecast verification community, the CRPS score has been found to produce sharper better-calibrated forecasts than with maximum likelihood estimation \cite{gebetsberger2018estimation}. 
\end{remark}

\begin{remark}[SVD]\label{remark:SVD:notes}
The SVD of the real $m \times n$ matrix $M$ is a decomposition into two orthogonal matrices $U,V$ and a diagonal matrix $\Sigma$, such that $M = U \Sigma V^T$, for $U, \Sigma, V^T \in \mathbb{R}^{m\times m},\mathbb{R}^{m\times n},\mathbb{R}^{n\times n}$. The diagonal entries $\sigma_{i}=\Sigma_{ii}$ of the diagonal matrix $\Sigma \in \mathbb{R}^{m\times n}$, are known as the singular values of $M$, they are uniquely defined by $M$ up to ordering and there are $rank(M)$ of them.
$U,V$ are orthonormal (rotation) matrices, interpreted as forming an orthonormal basis from columns or rows. The SVD is typically ordered such that the singular values are decreasing in size, while the matrix $\Sigma$ is unique in this ordering, matrices $U$ and $V$ are not unique. The columns of $U$ and the columns of $V$(rows of $V^T$) are called the left-singular vectors and right-singular vectors of $M$. A left singular value of $M$ is a non-negative real number $\sigma$, such that $
M v = \sigma u$ for $v\in \mathbb{R}^{n}$,$u\in \mathbb{R}^{m}$, a right singular value of $M$ as a non-negative real number $\sigma$, such that 
$M^T v = \sigma v$ for $v\in \mathbb{R}^{n}$,$u\in \mathbb{R}^{m}$.
We are interested in a truncated SVD, where we retain only the first $t$ singular values along with their corresponding $t$ column vectors in $U$, and $t$ row vectors in $V^{T}$. Such that $M_t=U_t\Sigma_t V^{T}_t$, for $U_t, \Sigma_t, V^T_{t} \in \mathbb{R}^{m\times t},\mathbb{R}^{t\times t},\mathbb{R}^{t\times n}$ approximates $M$, we use the Truncated SVD algorithm in Halko \cite{halko2011finding}.
\end{remark}

\begin{remark}[Fourier interpolation]\label{remark:fourier interpolation:notes}\label{remark:cost of interpolation:notes} Fourier interpolation of an unstructured grid of points using a structured 2d regular grid of data on $[0,1]\times [0,1]$ (weather stations) can be performed in the following manner. Let $q\in \mathbb{R}^{n_{c}\times n_{c}}$, denote the discrete field at $\wedge_d$, where $n_{c}$ is even, and $\hat{q} = \mathcal{F}_x\mathcal{F}_y(q)$ its discrete 2d Fourier transform. Then $q(x_l,y_l) = \operatorname{Re}( \sum_{-n_{c}/2\leq j,k \leq n_{c}/2} \hat{q} \exp(2\pi i j x_{l}/ \Delta x) \exp(2\pi i k y_{l} /\Delta y) )$, $\forall l$, is Fourier interpolation at $(x_l,y_l)$. This is performed on the x, and y components of (deterministic and stochastic) velocity fields, at discrete time evolving points $\b x_{l} \in \wedge_0(t)$ in the methods \cref{method:with} \cref{method:without}. Typically the error of Fourier interpolation is related to the continuity of the field. The reconstruction and evaluation of interpolating polynomials or other basis functions from weather stations has an inherent error and computational cost. Two dimensional Fourier interpolation from $n_c\times n_c$ weather stations to $n_v$ points has roughly the following computational cost, in floating point operations $C P n_c^4 2 \log(n_c) n_v $.
\end{remark}


 \begin{table}
    \centering
\begin{tabular}{||c c c c||} 
 \hline
Moment & dataset 1 & dataset 2 & expected value  \\
 \hline\hline
 $\mathbb{E}( \Delta W_r)$ & -6.349e-16 & -2.325e-16 & 0 \\
 \hline
$\mathbb{E}( \Delta W_r^2)/ \Delta t$ & 1.000e+00 & 1.000e+00  & 1 \\
\hline
$\mathbb{E}( \Delta W_r^3)$ & 8.875e-04 & -1.478e-03 & 0 \\
\hline
$\mathbb{E}( \Delta W_r^4)/ \Delta t^2 $& 2.727e+00 & 2.878e+00&3 \\
\hline
$\mathbb{E}( \Delta W_r^5)$ & -5.641e-03 &-2.375e-03& 0 \\
\hline
$\mathbb{E}( \Delta W_r^6)/ \Delta t^3$ & 1.224e+01 & 1.336e+01 & 15 \\ \hline 
\end{tabular}
\caption{Moments of the recovered increments.  }
\label{table: moments}
\end{table}

\begin{figure}[hbt!]
    \centering
\begin{subfigure}[t]{0.32\textwidth}
\centering
\includegraphics[width=.95\textwidth]{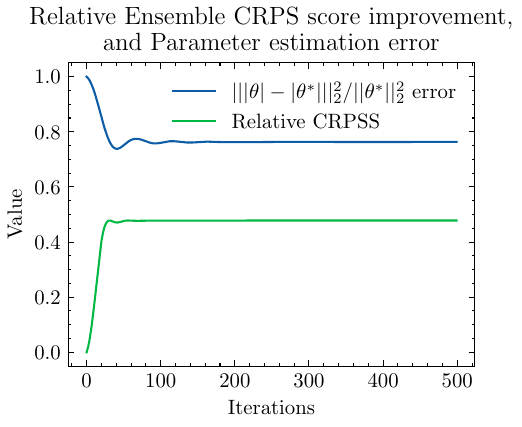}
\end{subfigure}
\begin{subfigure}[t]{0.32\textwidth}
\centering
\includegraphics[width=.95\textwidth]{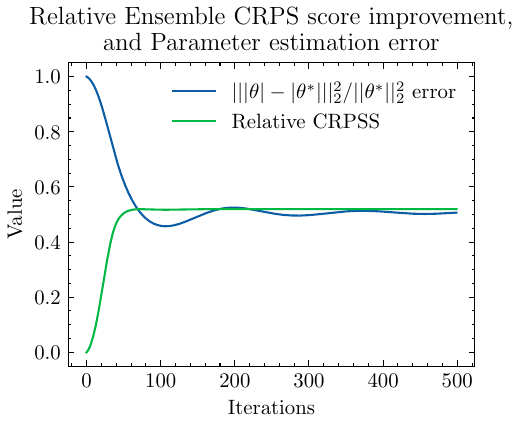}
\end{subfigure}
\begin{subfigure}[t]{0.32\textwidth}
\centering
\includegraphics[width=.95\textwidth]{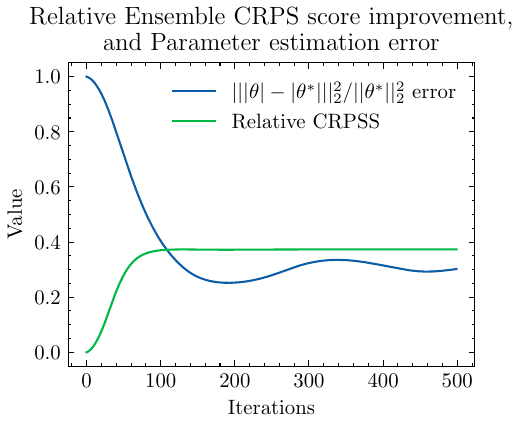}
\end{subfigure}
\caption{On a different dataset generated with fewer dimensions (using much less points), we were able to run the ensemble training over the full 256 time intervals, with 1,10,100 ensemble members respectively. The effect of increasing the ensemble size appeared to improve the parameter estimation error.  }
\label{fig:training on more ensemble members.}
\end{figure}

\begin{remark}[Notation, Initialisation and Weather stations]\label{remark:notation} 
Let $\wedge^{m,n}_{[x_{\min},x_{\max}]\times [y_{\min},y_{\max}]} = X^{m}_{[x_{\min},x_{\max}]} \times  Y^{n}_{[y_{\min},y_{\max}]} $, denote the $m \times n$ Cartesian product mesh defined from the following two sets $X^{m}_{[x_{\min},x_{\max}]}:=\lbrace x_i | x_i = x_{\min} + h_x (i+1/2), \forall i \in  \lbrace 0,...,m-1 \rbrace \rbrace$, $h_x := (x_{\max}-x_{\min})/m$,
$Y^{n}_{[y_{\min},y_{\max}]}:=\left\lbrace y_j | y_j = y_{\min} + h_y (j+1/2), \forall j \in  \lbrace 0,...,n-1 \rbrace \right\rbrace $, $h_y := (y_{\max}-y_{\min})/n$ of equally spaced cell center points. Such that the $(i,j)$-th element $(x_{i,j},y_{i,j})=(x_{i},y_{j})\in \wedge^{m,n}_{[x_{\min},x_{\max}]\times [y_{\min},y_{\max}]}$, denotes the $(i,j)$-th cell center position of a Cartesian product mesh on $[x_{\min},x_{\max}]\times[y_{\min},y_{\max}]$. We denote the vectorisation of an arbitrary $A\in \mathbb{R}^{m\times n}$ matrix by $$\operatorname{vec}(A)=\left[a_{1,1}, \ldots, a_{m, 1},a_{1,2}, \ldots, a_{m, 2},\ldots, a_{1, n}, \ldots, a_{m, n}\right]^{\mathrm{T}}\in \mathbb{R}^{nm\times 1}.$$  We denote the outer product by $\otimes$. We denote the vector of ones length $m$ by $\b e_m$. We denote the set of natural numbers from $1$ to $M$ by $[M]:=\lbrace 1,...,M \rbrace$. We denote the vector $\b x_{m}\in \mathbb{R}^{m}$, as the vector of increasing points in $X^{m}_{[x_{\min},x_{\max}]}$, and $\b y_{n}\in \mathbb{R}^{n}$ as the vector of increasing points in $Y^{n}_{[y_{\min},y_{\max}]}$. Such that we can construct coordinate matrices in the following manner $x_{i,j}= (\b e_{n} \otimes \b x_m)\in \mathbb{R}^{n\times m}$,  $y_{i,j} =(\b y_n \otimes \b e_m)\in \mathbb{R}^{n\times m}$. And evaluate the initial vorticity $\omega$ as a function of coordinate matrices in the following manner
$\Gamma_{i,j}=\omega(x_{i,j},x_{i,j})h_{x}h_{y}$, $\forall (i,j) \in [m]\times [n]$. We consider a $n \times n $ uniform Cartesian meshgrid $\wedge^{n,n}_{[x_{\min},x_{\max}]\times [y_{\min},y_{\max}]}$, of initial points in a closed subdomain of $\mathbb{R}^2$. With initial condition $\omega_{0}(x,y)$, evaluated at the positions of this uniform mesh such that $\Gamma_{i,j} = \omega_0(x_i,y_j)h_x h_y$ $\forall (i,j)\in \lbrace 1,...,n\rbrace \times \lbrace 1,...,m\rbrace$.  We define the following set of dynamically evolving points $\wedge_{0}:=\lbrace (x_{i,j},y_{i,j}) |  (X_{i,j},X_{i,j}) = (X_i,X_j)\in \wedge^{m,n},  \omega_0(X_i,X_j)\neq 0 \rbrace$, as those with nonzero initial vorticity on the initial mesh,
assumed $n_{v}\leq nm$ dimensional. Note that we have defined the initial condition mesh of points on the dual mesh to that in \cite{beale1985high,majda2002vorticity}. We consider a $n_c \times n_c $ Cartesian meshgrid $\wedge_{\b d} := \wedge^{n_c,n_c}_{[x_{\min,d},x_{{\max,d}}]\times [y_{\min,d},y_{{\max,d}}]}$, of fixed weather stations in a closed subdomain of $\mathbb{R}^2$ for all time. Where the subscript $d$ denotes ``data", and indicates that this is a weather-station position in which (velocity) data can be collected. 
\end{remark}
   
\end{appendices}

\end{document}